\documentclass[sigplan, screen]{acmart}
\setcopyright{cc}
\setcctype{by}
\acmDOI{10.1145/3774934.3786429}
\acmYear{2026}
\copyrightyear{2026}
\acmISBN{979-8-4007-2310-0/2026/01}
\acmConference[PPoPP '26]{Proceedings of the 31st ACM SIGPLAN Annual Symposium on Principles and Practice of Parallel Programming}{January 31 -- February 4, 2026}{Sydney, NSW, Australia}
\acmBooktitle{Proceedings of the 31st ACM SIGPLAN Annual Symposium on Principles and Practice of Parallel Programming (PPoPP '26), January 31 -- February 4, 2026, Sydney, NSW, Australia}
\received{2025-08-23}
\received[accepted]{2025-11-10}

\usepackage{graphicx}
\usepackage{subcaption}
\usepackage{amsmath}
\usepackage{array}
\usepackage{bbding}
\usepackage{makecell}
\usepackage{threeparttable}
\usepackage{booktabs}
\usepackage{multirow}
\usepackage{colortbl}
\usepackage{enumitem}
\usepackage{siunitx}

\usepackage[linesnumbered,ruled,vlined]{algorithm2e}
\usepackage{listings}
\usepackage{xcolor}

\usepackage{amsfonts}
\usepackage{algorithmic}
\usepackage{textcomp}
\usepackage{tagging}
\usepackage[most]{tcolorbox}
\usepackage{tikz}

\newcolumntype{L}[1]{>{\raggedright\arraybackslash}p{#1}}

\makeatletter
\renewcommand{\@authorfont}{\normalsize\linespread{0.98}\selectfont}
\makeatother

\begin{document}

\title[CCL-D]{CCL-D: A High-Precision Diagnostic System for Slow and Hang Anomalies in Large-Scale Model Training}         


\author{Yida Gu}
\authornote{Yida Gu and Fakang Wang contributed equally. }
\affiliation{
    \institution{University of Chinese Academy of Sciences}
    \city{Beijing}
  \country{China}
}
\email{guyida21@mails.ucas.ac.cn}

\author{Fakang Wang}
\authornotemark[1]
\affiliation{
  \institution{Ant Group}
  \city{Hangzhou}
  \country{China}
}
\email{fakang.wfk@antgroup.com}

\author{Jianhao Fu}
\affiliation{%
  \institution{Ant Group}
  \city{Hangzhou}
  \country{China}
}
\email{fujianhao.fjh@antgroup.com}

\author{Zhenhang Sun}
\affiliation{%
  \institution{Ant Group}
  \city{Hangzhou}
  \country{China}
}
\email{zhenhang.szh@antgroup.com}

\author{Qianyu Zhang}
\affiliation{%
  \institution{Ant Group}
  \city{Hangzhou}
  \country{China}
}
\email{qiongyu.zqy@antgroup.com}

\author{Hairui Zhao}
\affiliation{%
  \institution{Jilin University}
  \city{Changchun}
  \country{China}
}
\email{zhaohr21@mails.jlu.edu.cn}

\author{Xingchen Liu}
\affiliation{
  \institution{University of Chinese Academy of Sciences}
    \city{Beijing}
  \country{China}
}
\email{liuxingchen232@mails.ucas.ac.cn}

\author{Yang Tian}
\affiliation{%
  \institution{Ant Group}
  \city{Hangzhou}
  \country{China}
}
\email{lieyuan@antgroup.com}

\author{Wenjing Huang}
\affiliation{
  \institution{University of Chinese Academy of Sciences}
    \city{Beijing}
  \country{China}
}
\email{huangwenjing23@mails.ucas.ac.cn}

\author{Zedong Liu}
\affiliation{
  \institution{University of Chinese Academy of Sciences}
    \city{Beijing}
  \country{China}
}
\email{captain.liu77@gmail.com}

\author{Yifan Chen}
\affiliation{%
  \institution{Ant Group}
  \city{Hangzhou}
  \country{China}
}
\email{wugou.cyf@antgroup.com}

\author{Jinwu Yang}
\affiliation{
  \institution{University of Chinese Academy of Sciences}
    \city{Beijing}
  \country{China}
}
\email{yangjinwu24@mails.ucas.ac.cn}

\author{Yueyuan Zhou}
\affiliation{
  \institution{University of Chinese Academy of Sciences}
    \city{Beijing}
  \country{China}
}
\email{zhouyueyuan24@mails.ucas.ac.cn}

\author{Qian Zhao}
\affiliation{%
  \institution{Ant Group}
  \city{Hangzhou}
  \country{China}
}
\email{zq317110@antgroup.com}

\author{Haoxu Li}
\affiliation{
  \institution{University of Chinese Academy of Sciences}
    \city{Beijing}
  \country{China}
}
\email{lihaoxu21@mails.ucas.ac.cn}

\author{Tao Wang}
\authornote{Corresponding authors: Dingwen Tao and Tao Wang.}
\affiliation{%
  \institution{Ant Group}
  \city{Hangzhou}
  \country{China}
}
\email{junchen.wt@antgroup.com}

\author{Feng Yu}
\affiliation{%
  \institution{Ant Group}
  \city{Hangzhou}
  \country{China}
}
\email{chuba.yf@antgroup.com}

\author{Zhan Wang}
\affiliation{%
  \institution{University of Chinese Academy of Sciences}
  \city{Beijing}
  \country{China}
}
\email{wangzhan@cqxyy.net}

\author{Guangming Tan}
\affiliation{%
  \institution{University of Chinese Academy of Sciences}
  \city{Beijing}
  \country{China}
}
\email{tanguangming@cqxyy.net}

\author{Dingwen Tao}
\authornotemark[2]
\affiliation{%
  \institution{University of Chinese Academy of Sciences}
  \city{Beijing}
  \country{China}
}
\email{taodingwen@cqxyy.net}

\renewcommand{\shortauthors}{Gu and Wang et al.}

\begin{abstract}
As training scales grow, collective communication libraries (CCL) increasingly face anomalies arising from complex interactions among hardware, software, and environmental factors. These anomalies typically manifest as slow/hang communication, the most frequent and time-consuming category to diagnose. However, traditional diagnostic methods remain inaccurate and inefficient, frequently requiring hours or even days for root cause analysis. To address this, we propose CCL-D, a high-precision diagnostic system designed to detect and locate slow/hang anomalies in large-scale distributed training. CCL-D integrates a rank-level real-time probe with an intelligent decision analyzer. The probe measures cross-layer anomaly metrics using a lightweight distributed tracing framework to monitor communication traffic. The analyzer performs automated anomaly detection and root-cause location, precisely identifying the faulty GPU rank. Deployed on a 4,000-GPU cluster over one year, CCL-D achieved near-complete coverage of known slow/hang anomalies and pinpointed affected ranks within 6 minutes—substantially outperforming existing solutions.
\end{abstract}



\maketitle

\section{Introduction}

Deep learning is reshaping technical paradigms in critical domains, including but not limited to natural language processing~\cite{kenton2019bert,vaswani2017attention}, autonomous driving~\cite{zhao2024autonomous,cui2024survey}, and audiovisual technologies~\cite{lin2023vision}. Guided by scaling laws, model performance has been tightly linked to the number of parameters, driving rapid expansion in model size to unlock greater capabilities~\cite{kaplan2020scaling}. However, this aggressive scaling has outpaced improvements in system reliability, exposing large-scale distributed training to increasing failure risks~\cite{wang2023gemini}.

As model scales surpass tens of billions of parameters, ensuring training reliability has become both critical and increasingly challenging. For example, Llama3-405B experienced 419 failures over 54 days of training on 16,000 GPUs~\cite{dubey2024llama}, yielding a Mean Time Between Failures (MTBF) of just 3 hours. 
The underlying causes stem from two systemic factors~\cite{dong2024boosting,wang2023gemini,jiang2024megascale,dai2025ft}: 
(1) hardware failure rates grow linearly with device count, threatening training continuity; and 
(2) system complexity increases non-linearly with the scale of computation and communication. 
Together, these forces drive a superlinear rise in training failures, posing formidable obstacles that existing mitigation strategies struggle to overcome.


\textit{\textbf{Motivation.}}
While considerable efforts have focused on fault tolerance mechanisms—such as checkpointing~\cite{maurya2024datastates, wang2023gemini, eisenman2022check, li2024excp}—these solutions primarily aim to reduce recovery latency. However, they fall short of addressing the root causes of the widening reliability gap in large-scale training. Simply restarting failed tasks without identifying and addressing the underlying issue, such as an anomalous node, can lead to repeated failures and prolonged training delays. Therefore, diagnosing and pinpointing the causes of failures is essential to prevent recurrence, highlighting the \textit{urgent need} for high-precision diagnostic systems in large-scale training.


Among the various functionalities of diagnostic systems for detecting training interruptions, non-obvious communication slowdowns or hangs (Slow/Hang) caused by anomalies are particularly problematic. Unlike explicit issues—such as training misconfiguration, which conventional debugging tools or log inspection typically reveal~\cite{nguyen2025systematic}—these implicit anomalies often stem from hardware, software, or environmental factors and frequently evade timely identification.

\begin{figure}[!t]
  \centering  
  \includegraphics[width=\linewidth]{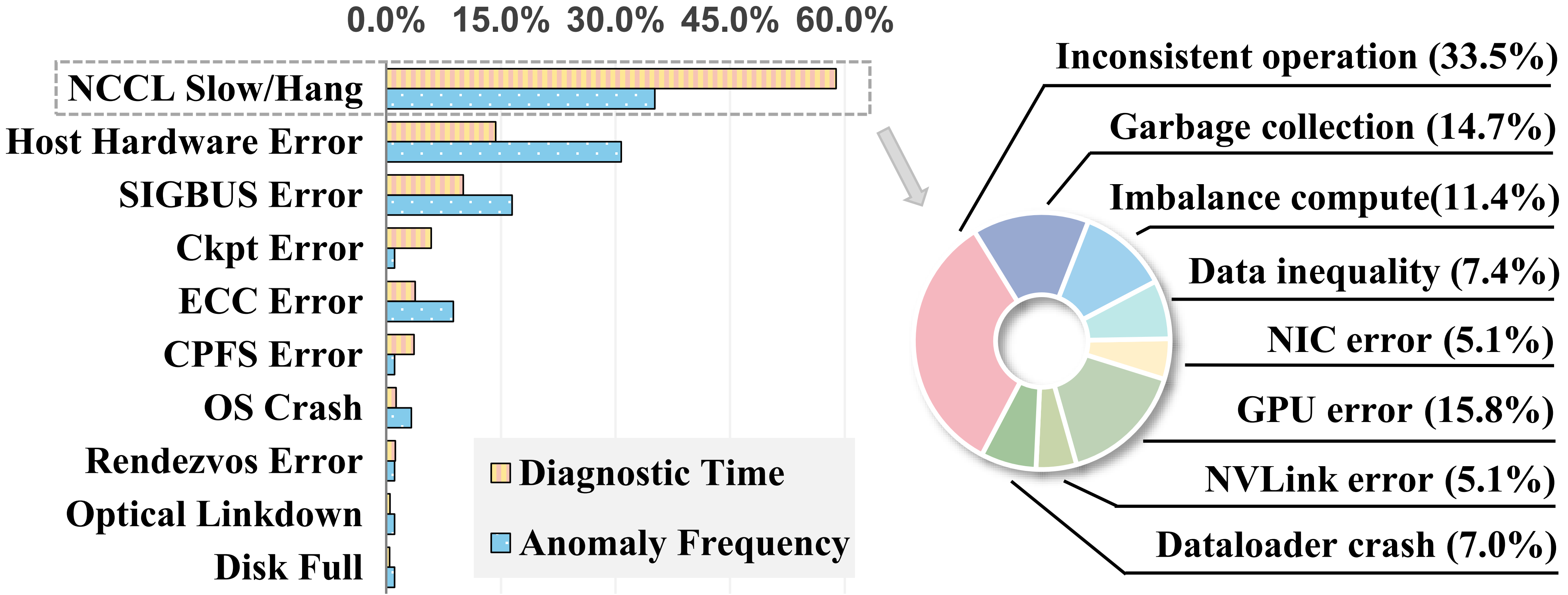}
  \caption{Training interruptions and slow/hang root-causes.}
  \label{fig:Introduction}
\end{figure}

For example, Figure~\ref{fig:Introduction} presents a 3-month observation from our 1,000-H800 GPU heterogeneous cluster running diverse model training tasks, during which 91 training interruptions occurred. Notably, slow and hang anomalies comprised 35.2\% of incidents yet consumed 58.8\% (70 hours) of total diagnostic time—making them the most frequent and costly to resolve. Their diverse and complex root causes (see Figure~\ref{fig:Introduction}, right) hinder effective diagnosis, and both their frequency and diagnostic overhead grow superlinearly with cluster size, highlighting the \textit{critical impact} and \textit{urgent need} for targeted mitigation measures.


\textit{\textbf{Limitations of Existing Solutions.}} Despite the availability of several diagnostic solutions, two limitations hinder their effectiveness in addressing slow/hang. \textbf{(1) Unacceptable low diagnostic accuracy.} Common mechanisms—such as PyTorch watchdog~\cite{watch_dog}, which triggers hang detection after a 30-minute timeout—offer no insights into the root cause. More importantly, 
most existing solutions~\cite{RAS, li2017parastack, gao2025dl2} largely overlook slow anomalies, and the few~\cite{wu2025greyhound, dong2024boosting} that address them lack the precision needed to pinpoint the exact location.
\textbf{(2) Prohibitive diagnostic overhead.} Constrained by their underlying diagnostic design, traditional methods—such as bisection-based stress testing~\cite{EDL} or expert-driven stack analysis~\cite{cui2025xputimer}—often take hours to days on large GPU clusters—rendering them impractical for real-time mitigation.
These gaps in diagnostic accuracy and efficiency have become a \textit{critical bottleneck} in advancing the reliability of large-scale distributed training systems.


\textit{\textbf{Our Solution.}} To address these challenges, we propose \textbf{CCL-D}
, a diagnostic system capable of automatically detecting and precisely locating slow/hang anomalies within minutes. The core idea behind CCL-D is to extend the functionality of CCL by enabling fine-grained identification and tracing of communication traffic. Rather than identifying the devices (e.g., switches or routers) that cause the anomalies, CCL-D aims to pinpoint the affected GPU(s) within the node topology and provide a probable root-cause analysis. To the best of our knowledge, \textit{CCL-D is the first diagnostic system to leverage both host-level and GPU kernel-level communication states for high-precision slow/hang diagnosis.}

Our main contributions are summarized as follows:
\begin{itemize}[leftmargin=1.em]
    \item We conduct a comprehensive analysis of collective communication workflows and slow/hang anomalies in different model training scenarios and, for the first time, derive six categories based on their potential root causes.
    
    \item We design a high-precision slow/hang diagnostic technique by extending CCL with a set of portable, cross-layer metrics centered on Send/Recv primitives. Integrated with a precision decision analysis algorithm, this design enables timely and accurate anomalies detection and location.

    \item We implement a lightweight distributed tracing framework that accurately captures and measures per-round communication traffic with negligible overhead (<1\%), enabling scalable deployment on production clusters.
    
    \item We validate CCL-D’s functionality and scalability by training large models on a 4,000-GPU cluster. Compared to the hours or even days required for model training, CCL-D diagnoses slow/hang anomalies with high precision within 6 minutes, significantly outperforming existing methods.
    
\end{itemize}

\section{Motivation and Challenges}

\subsection{Workflows of CCL Primitive Operations}
\label{ssec: ccl_workflow}

\begin{figure}[!t]
  \centering  \includegraphics[width=\linewidth]{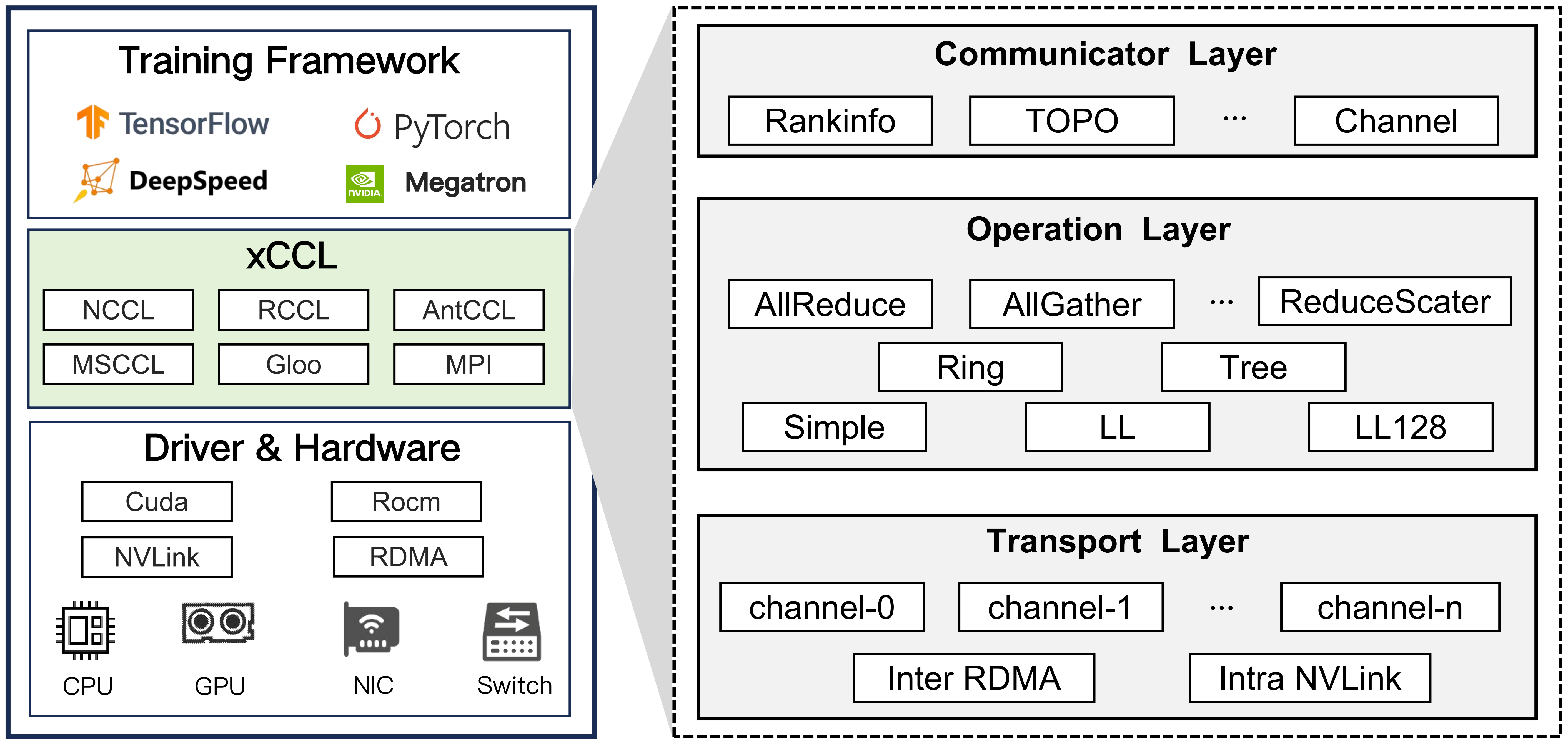}
  \caption{Position of CCL in training and its hierarchical structure.}
  \label{fig:CCL_bg}
\end{figure}

A single GPU can no longer meet the storage requirements of large models. To address this, distributed training frameworks utilize strategies such as DP~\cite{li2020pytorch, rajbhandari2020zero}, TP~\cite{wang2022tesseract}, and PP~\cite{huang2019gpipe,narayanan2019pipedream,zhao2025flexpipe,zhao2025arraypipe} to partition data and models, with each GPU storing a subset. This leads to the fact that training relies on collective communications to synchronize model parameters across GPUs. Therefore, as shown in Figure \ref{fig:CCL_bg}, the stability of CCL, which bridges high-level model services with low-level computational resources, is critical. 

In large-scale model training, different parallel strategies invoke various collective communication operations such as AllReduce, AllGather, ReduceScatter, and AlltoAll. Although their specific functions differ, these workflows can generally be abstracted into three critical phases: 1) Domain initialization: the host identifies participating ranks (i.e., GPUs) and establishes a communication domain, that is, the communicator; 2) Kernel dispatch: the CPU dispatching communication kernels (e.g., AllReduce) to the corresponding GPUs for execution; 3) Concurrent data transfer: multiple channels transfer data concurrently using intra-node NVLink~\cite{choquette2021nvidia} or inter-node RDMA~\cite{potluri2013efficient} hardware within kernels. Collective communications require all ranks within a communicator to behave consistently, including but not limited to algorithms (e.g., Ring, Tree~\cite{huang2021communication,cho2023logical}), protocols (e.g., Simple, LL, LL128~\cite{protocol}), and communication operation (i.e., GPU kernel) counts. Any deviation may lead to CCL slow/hang.

\subsection{Slow/Hang Anomalies Classification}
\label{ssec:Slow/Hang Anomalies Analysis}

Coarse-grained classifications such as "Slow" or "Hang" are insufficient for timely root cause identification and often delay remediation. Therefore, through analyzing slow/hang anomalies observed over the past 2 years in our production clusters and related reports in prior studies, we found that most issues are closely tied to communication workflows. Based on the analysis of Section \ref{ssec: ccl_workflow}, we derive six fine-grained anomalies' root causes (Figure \ref{fig:hang_slow}) that comprehensively cover slow/hang patterns in modern CCL systems: 

\begin{figure}[!t]
  \centering  
  \includegraphics[width=.95\linewidth]{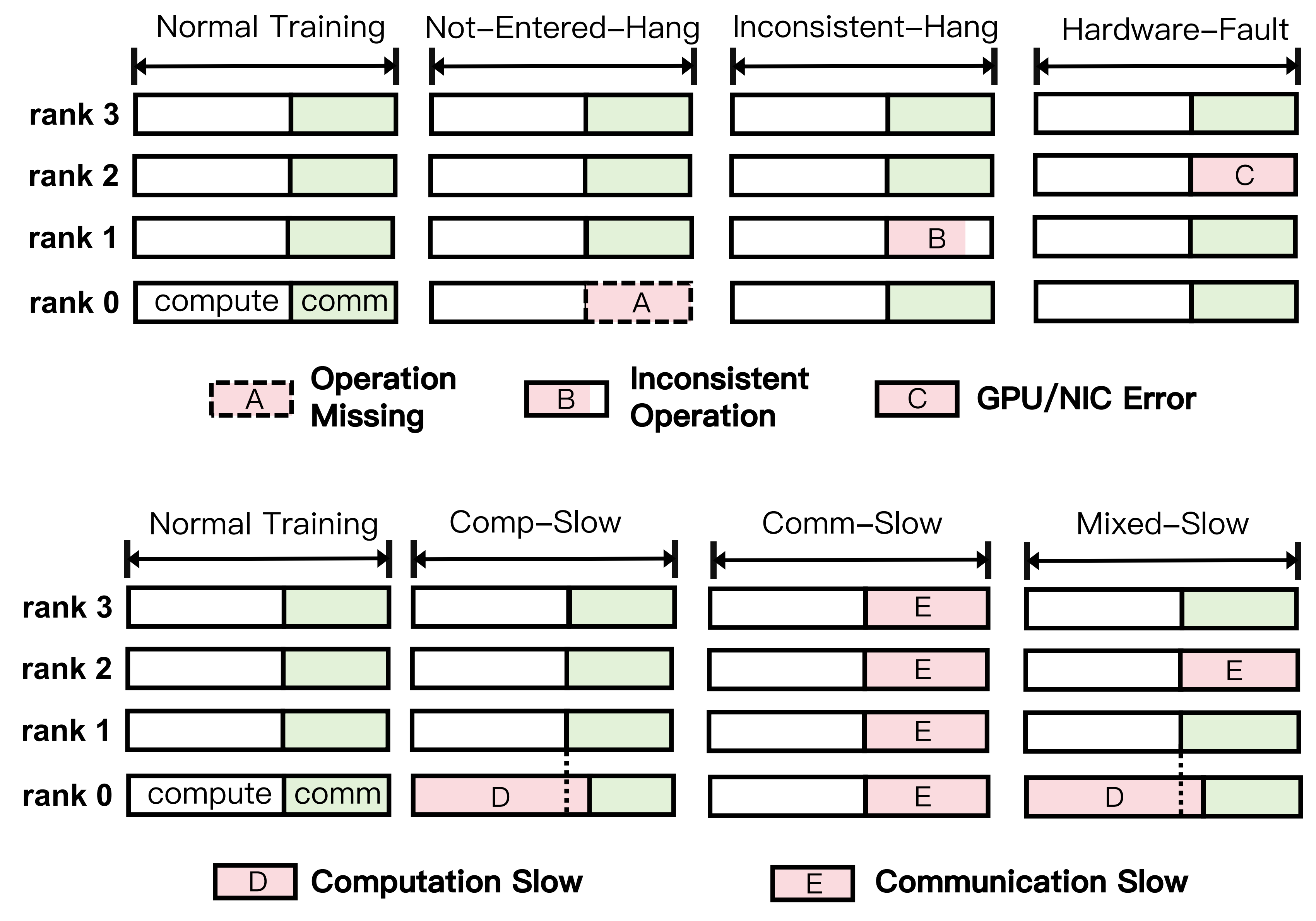}
  \caption{Hang (upper) and Slow (lower) scenarios analysis.}
  \label{fig:hang_slow}
\end{figure}

Hang Anomalies—\textbf{(H1) Not-Entered-Hang:} Some ranks in the communicator miss a communication operation and could not enter collective communication; \textbf{(H2) Inconsistent-Hang:} Inconsistent communication operations performed by individual GPUs within a communicator at the same time, leading to conflicts; and \textbf{(H3) Hardware-Fault:} Hardware (eg., GPU, NIC) or driver faults occurring on individual device during collective communication.

Slow Anomalies—\textbf{(S1) Computation-Slow:} Existing ranks enter communication later due to prolonged pre-computation, slower data loading, or GPU frequency throttling induced by thermal constraints, slowing down the communication;  \textbf{(S2) Communication-Slow:} Real-time network fluctuations or congestion can lead to performance degradation; and \textbf{(S3) Mixed-Slow:} Scenarios where both computation slow and communication slow occur simultaneously.

Over the past year, we have observed that the ratio of hang to slow cases was 62.1\% and 37.9\%, respectively. For hang, Not-Entered-Hang accounted for 11.8\%, Inconsistent-Hang for 58.9\%, and Hardware-Fault for 29.3\%. For slow, comp-slow represented 81.8\%, comm-slow 11.1\%, and mix-slow 7.1\%. Given the stability of collective communication workflows and hardware architectures in the near future, the proposed taxonomy remains broadly applicable and can be further refined for deployment-specific needs.


\subsection{Prior Diagnostic Works and Their Limitations}
\label{ssec:Communication Anomaly Detection}

Although several existing methods offer slow/hang anomaly detection or location, they each has specific limitations.

\textbf{Bisection-based Methods} are offline and reactive, relying on anomalies to be first detected manually before location can begin. When locating anomalies, tools like NCCL-tests~\cite{nccl-tests} are used to stress-test the involved machines and iteratively isolate faults. Training logs from OPT-175B~\cite{opt-log} and BLOOM-176B~\cite{bloom-log} show reliance on this method. However, due to the lack of runtime communication states, these methods only work for network- or hardware-related failures, and cannot reproduce logic-level or intermittent issues. Moreover, they are also time-consuming and resource-intensive, requiring task suspension during diagnosis.

\textbf{Stack Analysis} identifies anomalies by comparing function call information across all ranks. Although it can capture certain communication calls, stack traces are often verbose and complex, making their interpretation difficult and highly dependent on domain expertise and manual effort. In large-scale trainings, engineers must sift through the massive stack data generated by both CCL and the training components to make informed judgments. For example, ParaStack~\cite{li2017parastack} detects hangs by randomly sampling stack traces from selected communication processes and analyzing the specific states of MPI processes. Nevertheless, stack traces may fail to capture all anomaly types, particularly those involving loops or hardware slowdowns, leading to misdiagnosis.

\textbf{CCL Runtime Analysis} focuses on runtime communication states, traces behaviors of the collective communication during training. NCCL RAS~\cite{RAS} maintains a thread per rank to exchange state and monitor inconsistencies, but only offers coarse-grained information, such as current rank status and communication operation invocation counts, limiting its diagnostic resolution. Alibaba C4D~\cite{dong2024boosting} extends this approach with additional metrics (e.g., data transfer time, receiver wait time), enabling basic detection but low location precision of slow anomalies. Greyhound~\cite{wu2025greyhound} intercepts communication calls to monitor iteration time during training, detect slow anomalies only, halt training upon detection, and locate slow rank through stress testing. 

Unlike the first two categories, CCL-D represents the third class of approaches, diagnosing anomalies from CCL-level runtime states. By introducing fine-grained kernel metrics beyond host-level information, CCL-D achieves high-precision coverage of slow/hang anomalies with minimal overhead, while ensuring training continues uninterrupted.


\subsection{Our Intuition and Research Challenges}
\label{ssec:challenge}


\textit{Any slow/hang anomaly can be diagnosed by analyzing the CCL states}, which ultimately manifest as behavioral discrepancies among the participating ranks. This calls for finer-grained, multi-dimensional kernel-level metrics to more precisely capture communication behaviors and pinpoint anomalies. Nevertheless, two key challenges must be addressed:

\textit{\textbf{Challenge 1: Effective design of kernel metrics capable of accurately characterizing and diagnosing slow/hang anomalies remains a non-trivial task.}} 
Although existing profiling tools (e.g., NVIDIA NCU~\cite{NCU}) expose hundreds of runtime metrics (e.g., register utilization, SM occupancy), these signals are primarily designed to reflect the kernels' computation states rather than capturing the interactive nature of collective communication. Hence, they fail to reveal the root causes of slow/hang behaviors.
Furthermore, the complex many-to-many communication patterns and deeply nested loops inherent in collective operations make custom measurement logic, such as embedding timers, highly intrusive, requiring careful consideration of synchronization overhead and often altering kernel behavior. This results in high maintenance costs and poor portability. Therefore, it is crucial to identify kernel-level information tightly coupled with slow/hang anomalies and to construct a minimal, complementary subset to augment existing metrics.

\textit{\textbf{Challenge 2: Achieving high-precision communication traffic measurement and analysis under low overhead is a key challenge.}}
In large-scale training, hybrid parallelism introduces frequent and highly concurrent collective operations, with overlapping communicators that exacerbate measurement complexity. A naive solution, centralized registration and unified traffic management, would incur excessive synchronization delays and high-frequency data accesses, becoming itself a scalability bottleneck.
In addition, although directly measuring and storing kernel-level metrics on each rank is simpler to design, it would consume valuable GPU compute and memory resources, interfering with model training. Therefore, to prevent the diagnostic system from becoming a performance bottleneck, a lightweight and scalable metric tracking mechanism is essential.

\section{Overview of CCL-D}

\begin{figure}[!t]
  \centering  \includegraphics[width=\linewidth]{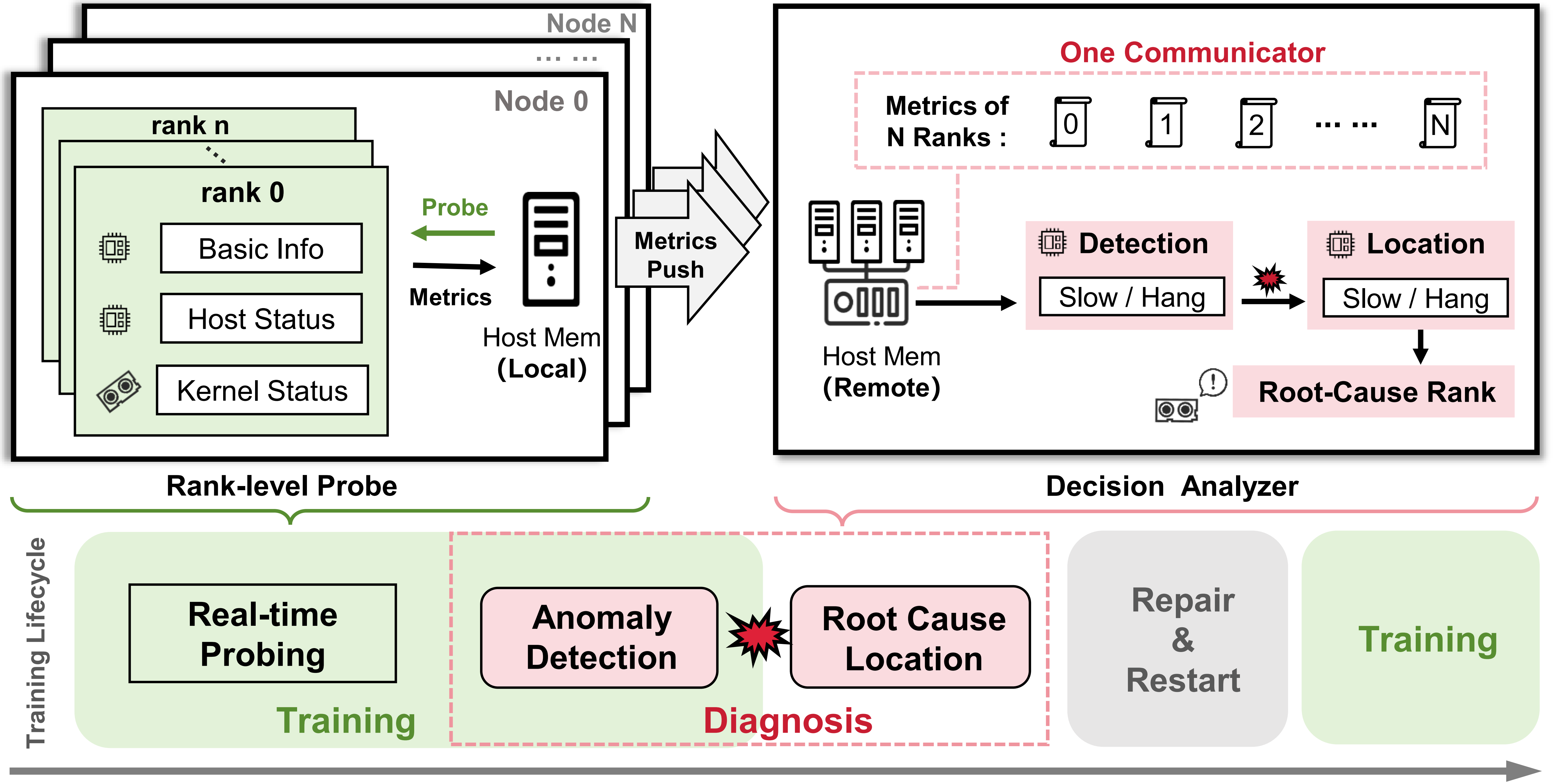}
  \caption{CCL-D and training lifecycle under anomalies.}
  \label{fig:overview}
\end{figure}

To address these challenges, we propose CCL-D, a diagnostic framework for rapid detection and location of slow/hang anomalies. As illustrated in Figure \ref{fig:overview}, CCL-D consists of two core modules: a distributed real-time rank probing module and a centralized decision analysis module.

The \textbf{Rank Probing Module} is deployed on each participating rank. It collects not only basic rank states but also a carefully designed set of multi-level metrics spanning both host and GPU kernel states to comprehensively characterize slow/hang anomalies (Section \ref{ssec:metric_selection}). \textit{Our design aims to minimize the number of metrics while maximizing their diagnostic coverage and utility, thereby reducing probing complexity.} A lightweight distributed tracing mechanism is implemented using basic rank information, and kernel metrics are measured with minimal overhead by leveraging host-side resources only (Section \ref{ssec:metric_collection}).

The \textbf{Decision Analysis Module} is a centralized but scalable component shared across ranks (Section \ref{ssec:decision_analyzer}). It periodically processes metrics from all ranks in a communicator to detect slow/hang anomalies, and upon detection, applies specialized decision algorithms to accurately locate the root-cause ranks. Unlike a single-node design, this module operates as a small distributed cluster for improved scalability.

Importantly, the decision analysis operates out-of-band, decoupling metric analysis from training execution to avoid any impact on training performance or accuracy. While online metric collection incurs a slight runtime overhead (<1\%), this impact is negligible compared to the significant reduction in the overhead of anomaly detection and location.

\section{Design of Metrics and Decision Analyzer}
\label{sec:design}

Before discussing metric measurement, this section first introduces the design of metrics, followed by how the decision analyzer leverages them for anomaly detection and location.


\subsection{Cross-layer Probing Metrics Design}
\label{ssec:metric_selection}

\begin{figure}[t]
  \centering  \includegraphics[width=\linewidth]{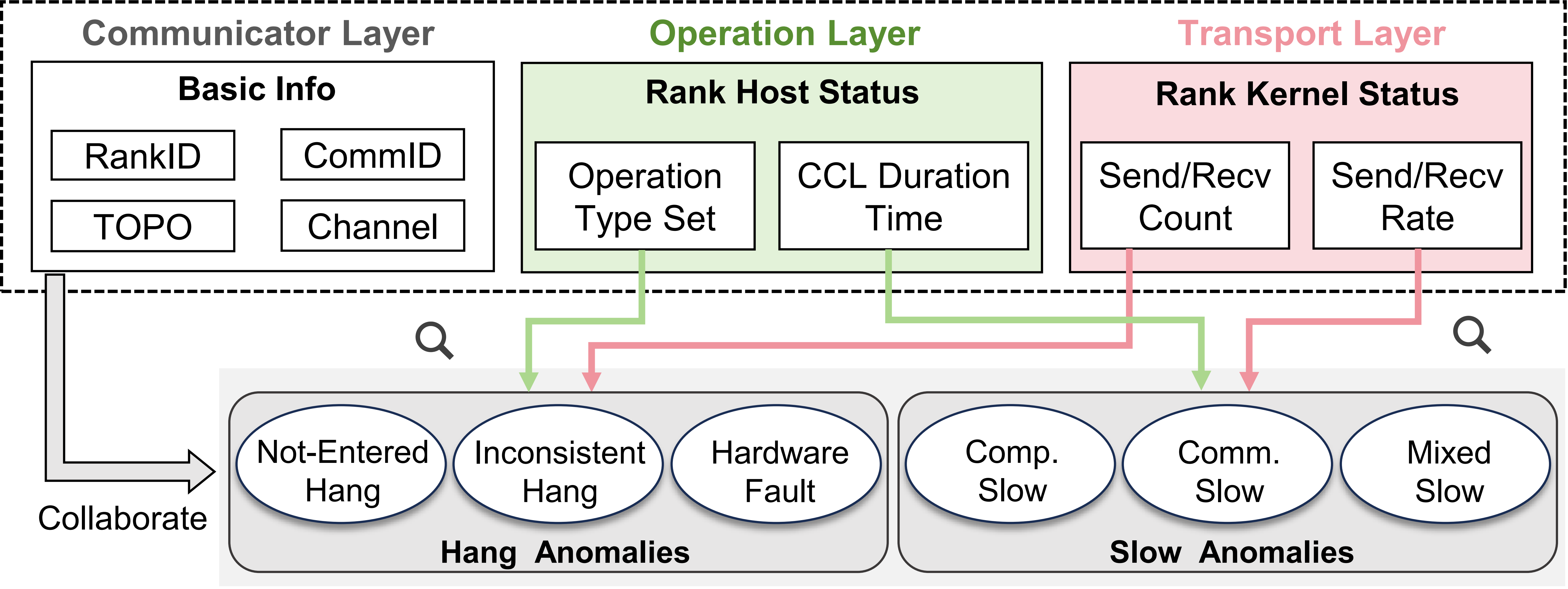}
  \caption{Metrics of CCL-D and corresponding anomaly types.}
  \label{fig:Metrics_selection}
\end{figure}

We design metrics across three layers of the CCL stack, providing multi-perspective coverage of communication behavior. Figure \ref{fig:Metrics_selection} shows the mapping between these metrics and the anomaly scenarios they target. The metrics in the basic information layer serve both as a means for communication traffic identification and as a foundation for basic diagnosis.

\subsubsection{Analysis of Kernel-level Metrics}

To improve diagnostic accuracy while minimizing system complexity, it is essential to identify metrics that capture the root causes of Hang/Slow anomalies, thereby reducing reliance on redundant kernel information. In the context of collective communication, which fundamentally consists of data exchange between GPUs, the atomic Send and Recv operations naturally meet these criteria. As the foundational primitives of all collective operations, they are independent of underlying hardware topologies, communication protocols, or CCL implementations, ensuring strong generality and portability. Empirical analysis further shows that in a correctly completed collective operation, all participating ranks exhibit consistent Send/Recv behaviors, demonstrating high comparability and structural regularity.

In contrast to methods that only monitor kernel invocations in the operation layer, Send/Recv activities provide fine-grained visibility into the transport layer and can expose low-level issues invisible at the host level. Moreover, by focusing solely on Send/Recv behavior, they introduce minimal overhead, require only lightweight kernel modifications, and are well-suited for scalable and efficient deployment.


\subsubsection{Metrics-Based Characterization of Slow/Hang}
\label{sssec:slow_hang_chara_metrics}

To tightly couple with the root causes of slow/hang, we design two types of kernel-level metrics based on Send/Recv behavior: one measuring count, the other rate.

\textbf{Hang Diagnostic Metrics.} Hang anomalies are fundamentally characterized by imbalanced Send/Recv counts across ranks. To capture this, we introduce \textbf{\textit{SendCount}} and \textbf{\textit{RecvCount}}, which record the actual number of send and receive instructions executed within the kernel. Compared to operation-layer invocation counts, these metrics can detect cases where operation counts are consistent but underlying Send/Recv behaviors diverge. To further support the detection of inconsistencies caused by mismatched communicator or scheduling errors, we also introduce a host-level metric—\textbf{\textit{Operation Type Set}}—that records static metadata for each rank, including the communication algorithm, protocol, data size, and operation name. These parameters remain constant throughout the entire communication.


\textbf{Slow Diagnostic Metrics.} Slow anomalies manifest as degraded Send/Recv rates. While operator-level timestamps can be used to estimate kernel \textbf{\textit{Duration Time}} per rank to make an initial judgment, this approach is limited by NTP clock drift~\cite{corbett2013spanner, marzullo1983maintaining} and the millisecond-to-microsecond scale of collective operations, making the time-based diagnosis inaccurate. Moreover, SendCount/RecvCount fail to capture such anomalies, as final counts may remain consistent despite underlying performance degradation.

\begin{figure}[t]
  \centering  
  \includegraphics[width=.97\linewidth]{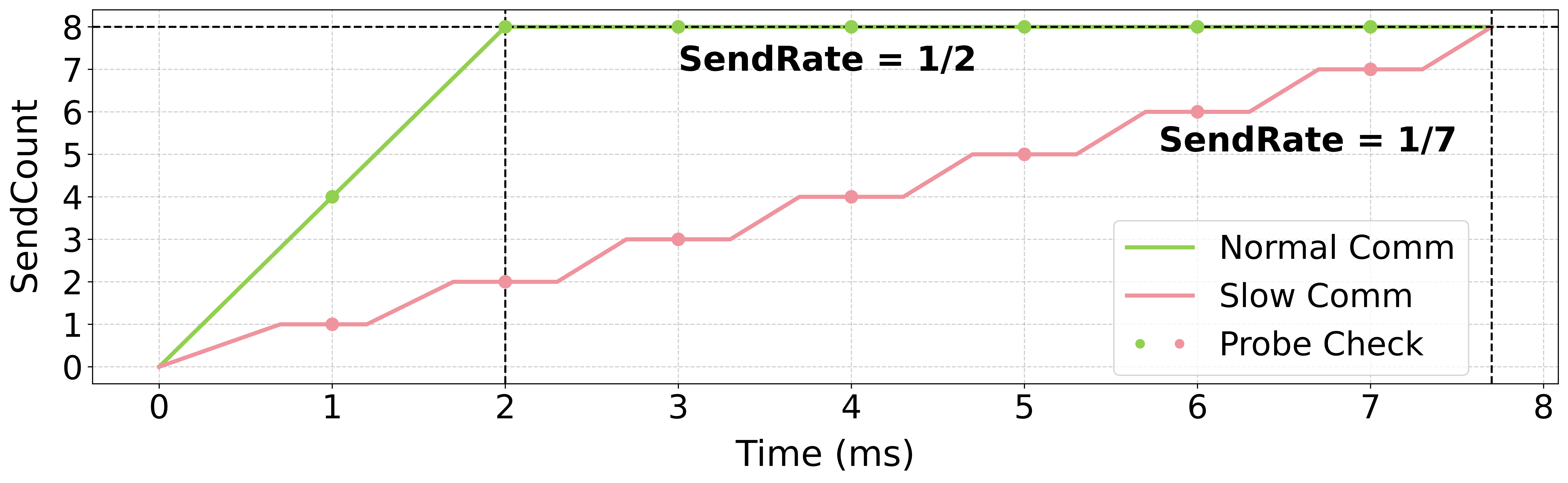}
  \caption{Comparison of SendRate between normal and slow ranks.}
  \label{fig:slow-change}
\end{figure}


To overcome this, we model communication progress as a cumulative count function $C(t)$. Under normal conditions, the derivative $dC/dt$ is nearly constant, reflecting a stable communication rate; during a slow anomaly, it drops markedly. We define \textbf{\textit{SendRate}} and \textbf{\textit{RecvRate}} as approximations of $dC/dt$, computed as the reciprocal of the number of changes in SendCount/RecvCount within a fixed sampling window. This design captures fine-grained rate variations without relying on global clock synchronization. Figure \ref{fig:slow-change} illustrates a specific use case, where the value change is probed every 1 ms. In normal communication, 8 send operations complete with 2 value changes, giving a rate of $1/2$; under a slow anomaly, the same operation takes 7 changes, dropping the rate to $1/7$ and clearly exposing the degradation.

\subsection{Architecture of Decision Analyzer}
\label{ssec:decision_analyzer}

Cross-layer metrics provide critical signals for diagnosing slow/hang anomalies but cannot alone determine their occurrence or pinpoint root causes. To address this, CCL-D employs a centralized decision analyzer that systematically aggregates metrics from all ranks for anomaly detection and root-cause location.


\subsubsection{Slow/Hang Automatic Detection}

To enable accurate anomaly detection, CCL-D uses the communicator ID to group metrics and applies specialized rules for analysis. Furthermore, our analyzer must distinguish between Ring and Tree communication algorithms to ensure correct grouping of comparison targets within the same communicator. While we do not focus on CCL-specific implementation details, we leverage the fundamental design principles of Ring and Tree to guide this grouping. In Ring, all ranks form a closed loop with equal in-degree and out-degree, enabling uniform analysis across the entire communicator. In contrast, Tree has a hierarchical structure in which only ranks within the same tree layer exhibit consistent SendCount/RecvCount and update frequencies; diagnostics are therefore restricted to same-layer ranks under tree topology.
 

\textbf{Hang anomaly detection} relies on the dimension of time. When the communication duration time of a rank exceeds the time threshold, our decision analyzer will trigger the creation of a hang anomaly alert. However, in large-scale model training, long-duration operations such as checkpoint or synchronization operations may occur normally. To reduce false positives, we set the hang threshold considering such long operations and further filter out alerts where the operation is \emph{AllReduce} with communication data size less than 4 Bytes (i.e., a barrier).

\textbf{Slow anomaly detection} is more nuanced and consists of two categories: \textit{slow-at-start} and \textit{in-communication slowdown}, each requiring a different temporal baseline. In the early training phase, when no historical statistics are available, the administrator specifies an initial baseline $T_{\text{base}}^{(\mathrm{init})}$ based on prior experience with similarly scaled models. The analyzer then dynamically updates this baseline by averaging the maximum durations of the first $m$ communication rounds, where $m$ is the smaller of 100 rounds or all rounds within the first two minutes, yielding $T_{\text{base}}^{(\mathrm{new})}$. Formally:
\begin{equation}
\small
T_{\text{base}} =
\begin{cases}
T_{\text{base}}^{(\mathrm{init})}, & \text{if } r \le m, \\[8pt]
\frac{1}{m} \sum\limits_{j=1}^{m} T_{\max}^{(j)}, & \text{otherwise}
\end{cases}
\end{equation}
where $r$ is the current number of communication rounds, and $T_{max}^{(j)}$ denotes the maximum duration time among all ranks participating in round $j$.

To capture representative slowdowns while avoiding the overhead of per-round analysis, we operate on a fixed one-minute detection window. Within each interval, the intra-round time range $\left(T_{\max}^{(r)} - T_{\min}^{(r)}\right)$ is computed for every round, and the round with the largest range is selected. Its maximum communication duration is defined as $T_{\max}$, focusing detection on the round with the most pronounced inter-rank disparity and avoiding misjudgments from relying solely on the absolute longest communication time. Formally:
\begin{equation}
\small
T_{\max} = T_{\max}^{\left( \arg\max\limits_{r \in \text{window}} \left[ T_{\max}^{(r)} - T_{\min}^{(r)} \right] \right)}
\end{equation}
Finally, the slowdown ratio is:
\begin{equation}
\small
\label{equation:slow_detection}
R = (T_{\max} - T_{\text{base}}) / {T_{\text{base}}}
\end{equation}


A slow anomaly is flagged when $R > \theta_{\mathrm{slow}}$. Instead of using a fixed subjective threshold, we determine $\theta_{\mathrm{slow}}$ statistically,
and in practice the threshold is usually close to 3. To further reduce false positives, cases involving barrier synchronization are excluded, and transient slowdowns caused by cluster jitter are ignored unless they recur. For this purpose, CCL-D maintains a cumulative counter of slow detections and triggers location only when the repetition threshold is exceeded.

\subsubsection{High-Precision Root Cause Location}

\begin{figure}[t]
  \centering  
  \includegraphics[width=\linewidth]{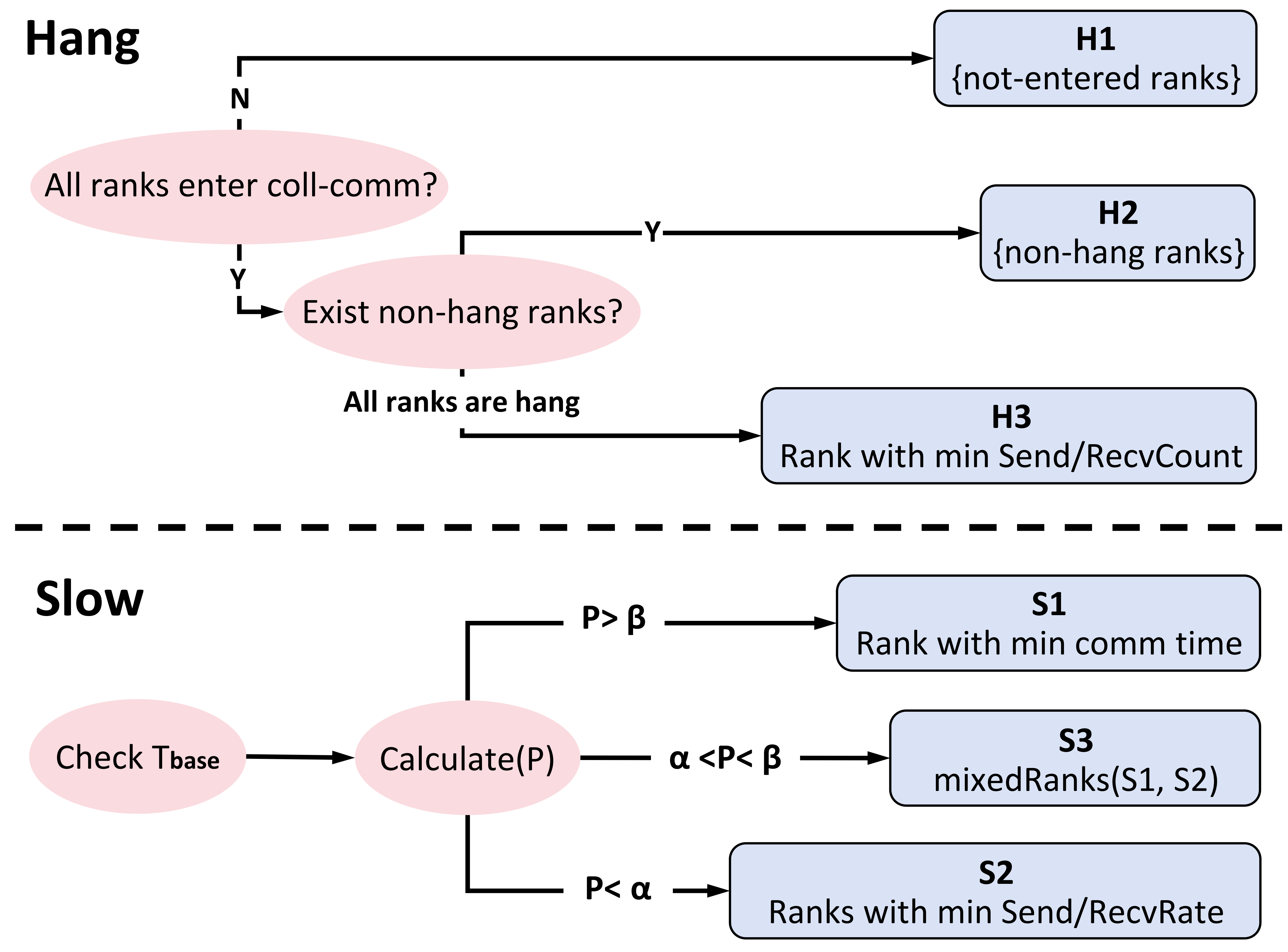}
  \caption{Decision tree of root-cause ranks. \textbf{H1-H3} and \textbf{S1-S3} correspond to the hang/slow discussed in Section~\ref{ssec:Slow/Hang Anomalies Analysis}.}
  \label{fig:decision-tree}
\end{figure}


Although detecting slow/hang can confirm anomalies within a communicator, it cannot directly identify the rank that triggered them. In large-scale clusters, anomalies may propagate across ranks, making the true root-cause rank difficult to isolate. To address this, the location module analyzes kernel metrics from all ranks and applies distinct algorithms to separate root-cause from secondary effects. Figure \ref{fig:decision-tree} illustrates the design logic. As the decision rules only compare metrics across participators, the algorithm runs in O(N) time for N ranks, remaining scalable to thousands of GPUs.

\textbf{Hang anomaly location} is formalized as a pure classification problem over rank states, using Trace ID as the first indicator (details in Section~\ref{ssec:data_structure}). A rank with no increment in its operation counter is inferred not to have entered the collective, and is identified as the root cause of a Not-Entered Hang. When all ranks have entered, two scenarios remain: One is the presence of non-hang ranks, which can be attributed to inconsistent operations, and the non-hang ranks are the root cause. The other is that all ranks are hang, suggesting a problem with hardware. The rank with the fewest SendCount/RecvCount is considered the root-cause rank.


\textbf{Slow anomaly location} begins by checking whether the current $T_{\text{base}}$ is a configured value. A static baseline suggests a slow-at-start scenario, while a dynamically updated baseline indicates in-communication slowdown, guiding subsequent handling strategies accordingly. We then introduce $P$ to formalize the relative contribution of computation delay versus communication delay in the current round:
\begin{equation}
\small
    P = (T_{max} - T_{min})/(T_{max} -T_{base})
    \label{equation:slow_location}
\end{equation}
$T_{min}$ and $T_{max}$ are similar in definition, except that the former takes the minimum communication time.  


In Equation (\ref{equation:slow_location}), $T_{\min}$ serves as a sliding indicator within the range $[T_{base}, T_{max}]$. When computation is the bottleneck, the last rank entering communication drives $T_{\min}$ close to $T_{base}$, making the numerator and denominator nearly equal and pushing $P \to 1$. Conversely, when communication dominates, $T_{\min}$ approaches $T_{max}$, diverging from $T_{base}$ and driving $P \to 0$. Conceptually, $P=0$ corresponds to purely slow communication, $P=1$ to purely slow computation, and $P=0.5$ to an equal contribution of both. However, in large-scale clusters, the equal-contribution scenario rarely occurs. To emphasize the dominant anomaly type, we introduce two boundary parameters $\alpha$ and $\beta$ around 0.5 (e.g., $\alpha=0.4$, $\beta=0.6$). When $P > \beta$, the anomaly converges to computation-slow, and the rank with minimal communication time is designated as the root cause; when $P < \alpha$, it is attributed to communication-slow, with the minimal SendRate/RecvRate rank as the root cause. Values in between suggest a mixed slowdown, requiring analysis of both slow types. This formalization provides a principled basis for rank attribution.

\section{Metric-Aware Tracing Framework}
\label{ssec:metric_collection}

Large-scale high-precision diagnostics face critical overhead challenges spanning metrics probing, storage, and submission. Unoptimized overhead impedes system scalability and hinders thousand-GPU model training. To overcome this, we design a lightweight distributed tracing framework that achieves precise communication traffic attribution and low-overhead metrics measurement via decentralized identification and host-driven measurement workflow.

\subsection{Distributed Communication Tracing}
\label{ssec:data_structure}\

To achieve precise and scalable recognition of concurrent communications, we introduce two complementary data structures: the Trace ID at the communicator level and the Probing Frame at the rank level. 

\begin{figure}[t]
  \centering  \includegraphics[width=0.7\linewidth]{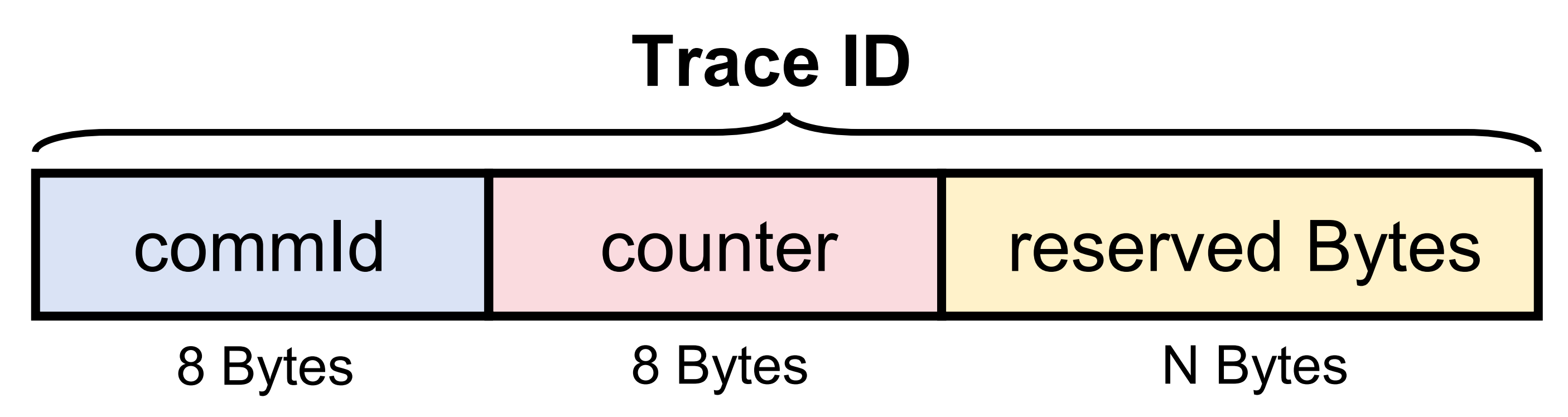}
  \caption{The structure of Trace ID.}
  \label{fig:traceid}
\end{figure}
 
\textbf{Trace ID.} This structure provides a decentralized mechanism to uniquely label each round of communication. As shown in Figure \ref{fig:traceid}, Trace ID consists of the communicator ID, an communication operation counter, and an optional extension field. Since all participating ranks increment their local counters synchronously at the start of each round, Trace ID ensures consistent operation labeling across the communicator. The extension field accommodates timestamps or status flags, supporting fine-grained traceability.

\begin{figure}[t]
  \centering  
  \includegraphics[width=.97\linewidth]{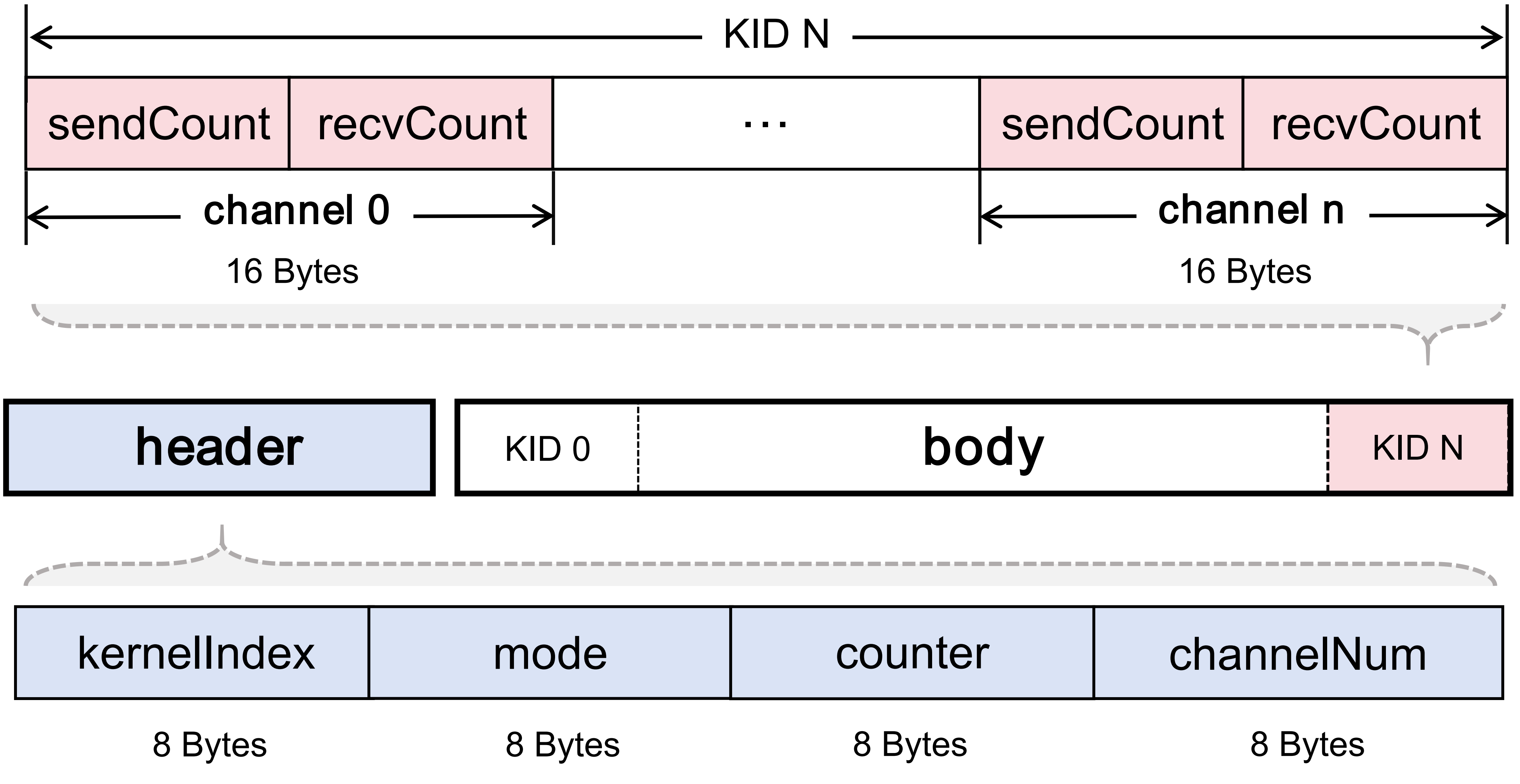}
  \caption{The structure of Probing Frame.}
  \label{fig:header_body}
\end{figure}

\textbf{Probing Frame.} Each rank maintains a Probing Frame to capture runtime states of communication kernels. As shown in Figure \ref{fig:header_body}, structurally consisting of a header and a body, it provides a compact and reusable mechanism for recording kernel-specific metrics. Since GPU kernels execute first-in-first-out (FIFO), we can avoid frequent memory allocation overhead by reusing the header and body. Hence, only one probing frame per rank is sufficient.


The header uniquely identifies each communication kernel and contains four fields: the operation counter, a mode flag (indicating whether metric measurement is enabled), the kernelIndex, and the number of communication channels. The body is cyclically partitioned into blocks, with kernelIndex specifying the block position for the current operation (computed as counter modulo the number of blocks). Within each round of communication, every channel is assigned two consecutive slots to record the SendCount and RecvCount. The number of channels that is correlated to the number of NICs and established during the CCL initialization. By separating records across channels, this design not only supports fine-grained diagnostic but also distinguishes intra-node RDMA anomalies from inter-node NVLink anomalies, since a rank may employ different channels in successive phases.

Together, these two diagnostic data structures enable lightweight multilevel tracing and storage of collective communication states, ensuring both consistency and scalability of cross-layer diagnostic information.


\subsection{Host-Driven Metrics Measurement}\
 
Directly measuring metrics on GPUs not only risks memory overflow, but also interferes with normal training execution. To avoid this, the measurement process must remain non-blocking and non-intrusive to GPU kernels. Our design principle is therefore to minimize GPU memory and compute overhead while preserving diagnostic accuracy. We further observe that host resources are typically underutilized during training, consistent with prior studies~\cite{wang2023reliable,jiang2020unified,jeon2019analysis}. Guided by this insight, CCL-D shifts most of the metric measurement and computation to the CPU and host memory.

\begin{figure}[t]
  \centering  
  \includegraphics[width=.95\linewidth]{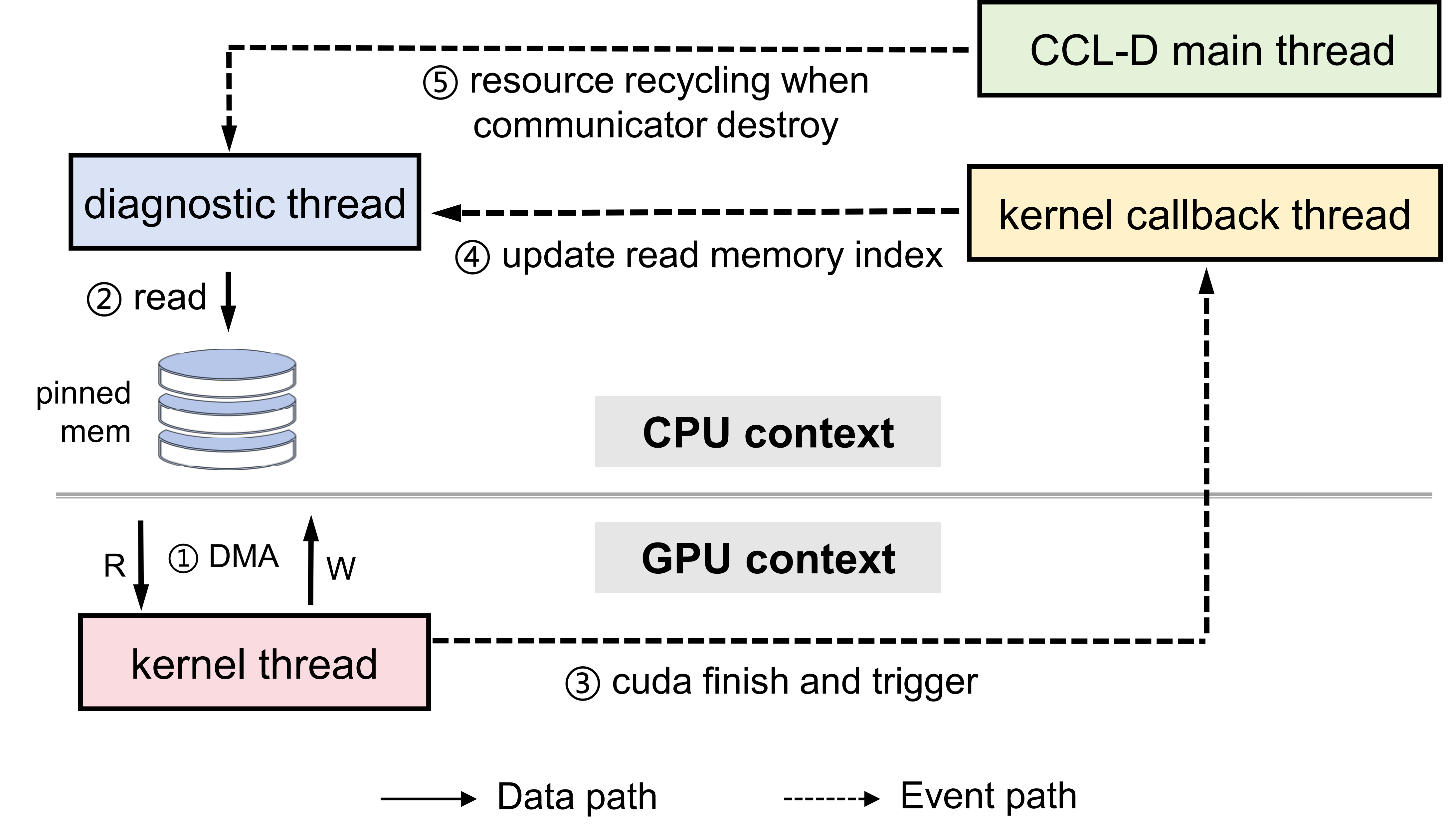}
  \caption{Metrics measurement workflow.}
  \label{fig:workflow}
\end{figure}

Specifically, CCL-D employs a zero-copy memory sharing strategy based on CUDA Unified Virtual Addressing (UVA), enabling implicit DMA propagation of metrics without explicit synchronization. This contiguous pinned memory shared between GPU and CPU stores the probing frames of all local ranks, allowing high-frequency metric updates during Send/Recv operations (\textcircled{1} in Figure~\ref{fig:workflow}). To further avoid occupying GPU computational resources, a CPU diagnostic thread periodically reads the pinned memory to perform lightweight computations, such as updating kernels' SendRate/RecvRate from their SendCount/RecvCount changes (\textcircled{2}). Upon kernel completion, a callback function is invoked (\textcircled{3}), which notifies the diagnostic thread to push current metrics to the decision analyzer and advance the buffer pointer to the next block (\textcircled{4}). Finally, once the communicator is destroyed, associated diagnostic resources are released promptly (\textcircled{5}), ensuring minimal overhead on host.


\section{Evaluation}

\subsection{Evaluation Setup}

\textbf{Platform:} We evaluate CCL-D on a GPU cluster where each node is equipped with 2 Intel 8469C CPUs, \SI{800}{GB} RAM, and 8 Nvidia H20 GPUs with \SI{96}{GB} HBM3 each, interconnected via \SI{900}{GB/s} NVLink. The nodes are further connected through 4 ConnectX-7 400G NICs. All nodes have the same software, including CUDA 12.2, NCCL 2.24.3,  Pytorch 2.4.0, and Megatron 0.9.0. For functional verification, we use two nodes with 16 GPUs. To assess scalability, we expand the experiments up to 4,000 GPUs. 
To show the generality of CCL-D, we also deploy and evaluate its ROCm-based counterpart, RCCL~\cite{RCCL}.

\begin{table*}[t]
    \caption{Comparison of diagnosis accuracy and efficiency across baseline methods and CCL-D.}
    \label{table:merged-diagnosis}
    \centering
    \renewcommand{\arraystretch}{1.1}
    \setlength{\tabcolsep}{9pt}
    \resizebox{\linewidth}{!}{
        \begin{tabular}{c|cccccc|cc|cc|c}
\toprule[0.5mm]
\multirow{3}{*}{\textbf{\makecell{Diagnosis\\Method}}} 
& \multicolumn{6}{c|}{\textbf{Diagnosis Accuracy}} 
& \multicolumn{4}{c|}{\textbf{Diagnosis Efficiency}} 
& \multirow{3}{*}{\textbf{\makecell{Runtime\\Dependency}}} \\
\cmidrule(lr){2-7} \cmidrule(lr){8-11}
& \makecell{Not-Entered\\Hang} 
& \makecell{Inconsistent\\Hang} 
& \makecell{Hardware\\Fault} 
& \makecell{Comp.\\Slow} 
& \makecell{Comm.\\Slow} 
& \makecell{Mixed\\Slow} 
& \makecell{Hang\\Detect} 
& \makecell{Slow\\Detect} 
& \makecell{Hang\\Locate} 
& \makecell{Slow\\Locate} 
& \\
\midrule[0.4mm]
Bisection 
& \color[HTML]{EF949E}\XSolidBrush 
& \color[HTML]{EF949E}\XSolidBrush 
& \color[HTML]{92D050}\Checkmark 
& \color[HTML]{EF949E}\XSolidBrush 
& \color[HTML]{92D050}\Checkmark\kern-1.3ex\raisebox{1ex}{\rotatebox[origin=t]{125}{\textbf{---}}} 
& \color[HTML]{EF949E}\XSolidBrush 
& $>30$\,min & $>1$\,h & 4\,min & 4\,min & Offline \\
Stack 
& \color[HTML]{92D050}\Checkmark 
& \color[HTML]{92D050}\Checkmark 
& \color[HTML]{92D050}\Checkmark 
& \color[HTML]{EF949E}\XSolidBrush 
& \color[HTML]{EF949E}\XSolidBrush 
& \color[HTML]{EF949E}\XSolidBrush 
& $>30$\,min & N/A & 5\,min & N/A & Online \\
RAS 
& \color[HTML]{92D050}\Checkmark 
& \color[HTML]{EF949E}\XSolidBrush 
& \color[HTML]{EF949E}\XSolidBrush 
& \color[HTML]{EF949E}\XSolidBrush 
& \color[HTML]{EF949E}\XSolidBrush 
& \color[HTML]{EF949E}\XSolidBrush 
& $>30$\,min & N/A & 10\,ms & N/A & Online \\
Greyhound 
& \color[HTML]{EF949E}\XSolidBrush 
& \color[HTML]{EF949E}\XSolidBrush 
& \color[HTML]{EF949E}\XSolidBrush 
& \color[HTML]{92D050}\Checkmark\kern-1.3ex\raisebox{1ex}{\rotatebox[origin=t]{125}{\textbf{---}}} 
& \color[HTML]{92D050}\Checkmark 
& \color[HTML]{EF949E}\XSolidBrush 
& N/A & 1min & N/A & 1.43\,s & Online  + Offline \\
C4D
& \color[HTML]{92D050}\Checkmark 
& \color[HTML]{EF949E}\XSolidBrush 
& \color[HTML]{EF949E}\XSolidBrush 
& \color[HTML]{EF949E}\XSolidBrush 
& \color[HTML]{92D050}\Checkmark 
& \color[HTML]{EF949E}\XSolidBrush 
& 5\,min & 1min & 104\,ms & 138\,ms & Online \\
\textbf{CCL-D} 
& \color[HTML]{92D050}\Checkmark 
& \color[HTML]{92D050}\Checkmark 
& \color[HTML]{92D050}\Checkmark 
& \color[HTML]{92D050}\Checkmark 
& \color[HTML]{92D050}\Checkmark 
& \color[HTML]{92D050}\Checkmark 
& \textbf{5\,min} & \textbf{1\,min} & \textbf{108\,ms} & \textbf{146\,ms} & \textbf{Online} \\
\bottomrule[0.5mm]
\end{tabular}
    }
    \begin{minipage}{0.94\linewidth}
        \vspace{1mm}
        \footnotesize
        \raggedright
        $\bullet$\ In all experiments, the communication algorithms (Ring, Tree) and protocols (Simple, LL, LL128) did not affect the results. \\
        $\bullet$\ N/A: The method does not support this anomaly diagnosis.
    \end{minipage}
    \vspace{-2mm}
\end{table*}

\textbf{Baselines:} We select 5 approaches detailed in Section \ref{ssec:Communication Anomaly Detection}. The bisection method works by using a divide-and-conquer approach and stress-testing with NCCL-tests. We use DLRover~\cite{EDL} as the specific implementation for bisection. For stack analysis, we develop our own baseline by integrating the design principles of XPUTimer~\cite{cui2025xputimer} and ParaStack. For CCL runtime analysis, we use NCCL RAS~\cite{RAS}, Greyhound~\cite{wu2025greyhound}, and C4D~\cite{dong2024boosting} as baselines for evaluation. 

\textbf{Parameters:} For fairness, both C4D and CCL-D use a 5-minute hang threshold and a 1-minute slow-detection window. 5-minute bound reflects checkpoint and normal barrier durations, and our cluster statistics show that 97\% of cases exceeding this limit cannot recover, making it a reliable cutoff. 1-minute  window not only complies with cluster Service Level Objectives (SLO), but also captures sufficient collective communication rounds to avoid per-round diagnosis overhead, thus balancing sensitivity and diagnostic.




\textbf{Workloads:} We use 2 communication algorithms, Ring-based and Tree-based~\cite{huang2021communication,cho2023logical}, and consider 3 communication protocols, Simple, LL and LL128~\cite{protocol}, to test the accuracy and efficiency of CCL-D. BaiLing-5B, Llama2-7B~\cite{touvron2023llama}, Llama3.1-8B~\cite{dubey2024llama} and BaiLing-80B~\cite{team2025every} are used as training tasks to measure and analyze the various overheads of CCL-D. We use Alpaca~\cite{alpaca} and Fineweb-edu~\cite{penedo2024fineweb} datasets to train them with two training strategies, FSDP~\cite{zhao2023pytorch} and 3D~\cite{shoeybi2019megatron} parallel.

\subsection{Performance of Anomaly Diagnostic}
\label{ssec:e_diagnosis_performance}

\subsubsection{Diagnostic Accuracy and Efficiency}

An effective diagnostic system must balance both accuracy and efficiency. Table \ref{table:merged-diagnosis} compares six diagnostic approaches from these two perspectives. This experiment is conducted on 16 devices. To comprehensively evaluate the detection and location capabilities, we construct diverse anomaly scenarios. For hang, cases include process blocking (e.g., SIGSTOP/SIGCONT), inconsistent operations, and NIC/GPU failures. For slow, we simulate computational bottlenecks (e.g., GPU frequency throttling, garbage collection interference), communication delays (e.g., link jitter, network misconfiguration, artificial delays), and system-level resource contention. This coverage ensures a thorough assessment of the diagnostic system’s effectiveness.

\textbf{Diagnostic Accuracy.} It encompasses two aspects: \textit{coverage of anomaly scenarios and precision of faulty-rank location}. By leveraging multi-dimensional metrics such as send/recv counts and rates, CCL-D achieves comprehensive coverage of both slow/hang anomalies with precise rank-level attribution. For RCCL diagnostic, CCL-D achieves the same diagnostic accuracy as NCCL scenarios.

In contrast, baseline methods show clear limitations. Bisection cannot reproduce real-time slow/hang and is restricted to hardware or network faults, requiring long testing cycles for location. Stack analysis covers all hang anomalies but fails to detect slow ones, since stack states remain identical across ranks. RAS only records host-level operation counts, which makes it incapable of handling slow anomalies and occasionally producing false positives in our experiments. Greyhound lacks mechanisms for hang; for comp-slow, it only captures computation delays caused by GPU frequency throttling, while failing to account for algorithm-level factors such as garbage collection. C4D lacks kernel-level insight and only relies on coarse-grained timing information, which makes it can only identify comm-slow in practice. Notably, both C4D and Greyhound fail to handle mixed-slow scenarios, leading to reduced location accuracy.


\textbf{Diagnostic Efficiency.} We evaluate efficiency in terms of anomaly detection latency and location latency, while also considering the impact of communication algorithms, protocols, and parallel strategies. Each test is repeated five times, and the average result is reported.

Bisection, stack analysis, and RAS do not automatically detect anomalies; instead, they rely on manual perception and intervention, making detection times highly variable. For fairness, we assume they operate under PyTorch Watchdog’s default 30-minute timeout, resulting in average hang detection delays of no less than 30 minutes. Based on production statistics, users typically perceive slow anomalies only after about 1 hour, which we adopt as the baseline for slow detection by bisection. Our experiments also confirm that communication algorithms, protocols, and parallel modes have negligible influence on CCL-D’s detection efficiency.

Regarding location efficiency, bisection requires multiple group tests to re-collect runtime information, while stack analysis depends on expert knowledge to compare stack differences—both inefficient, exceeding 4 minutes. RAS achieves lower latency, but since it relies solely on host-level operation counts, it can only identify Not-Entered-Hang, sacrificing both accuracy and coverage. Greyhound halts training and relies on stress testing, yielding lower efficiency than C4D and CCL-D, with second-level latency compared to their millisecond-level performance. Ultimately, CCL-D precisely locates slow/hang anomalies within \SI{150}{ms} under NCCL/RCCL, which is well suited for production-scale training.



\begin{table}[t]
  \caption{Comparing slow/hang diagnostic efficiency in one-month periods before and after CCL-D integration \textbf{(GPU scale: 16-4000)}.}
  \label{table:before-CCL-D}
  \resizebox{1\linewidth}{!}{
    \renewcommand{\arraystretch}{1.6} 

\begin{tabular}{@{}>{\centering\arraybackslash}m{1.2cm}
                >{\centering\arraybackslash}m{2.0cm}
                >{\centering\arraybackslash}m{2.0cm}
                >{\centering\arraybackslash}m{2.5cm}
                >{\centering\arraybackslash}m{3cm}@{}}
\toprule[0.5mm]
\textbf{Type} &
\makecell{\textbf{Diagnosis} \\ \textbf{Mode}} &
\makecell{\textbf{Cases} \\ \textbf{Captured}} &
\makecell{\textbf{Avg. Diagnosis} \\ \textbf{Time}} &
\textbf{Root Causes} \\
\midrule[0.4mm]
\multirow{2}{*}{Hang} & Manual & 4  & 47\,h   & NIC(H3), GPU(H3), UN \\
                      & \textbf{CCL-D}   & \textbf{18} & \textbf{6\,min (5+1)}  & \makecell{IO(H2), GPU(H3) \\ NV(H3), NE(H1)} \\
\midrule
\multirow{2}{*}{Slow} & Manual & 4  & 74\,h   & UN \\
                      & \textbf{CCL-D}   & \textbf{10} & \textbf{2\,min (1+1)}  & \makecell{DI(S1), GC(S1) \\ IC(S1), MS(S3)} \\
\bottomrule[0.5mm]
\end{tabular}
  }
  \begin{minipage}{\linewidth}
    \vspace{1mm}
    \footnotesize
    \raggedright
    $\bullet$\ \textbf{Abbreviations:} UN=Unknown; NIC/GPU=NIC/GPU error; NV=NVLink error; IO=Inconsistent operation; NE=Not entered; DI=Data inequality; GC=Garbage Collection; IC=Imbalance Compute; MS=Mixed-Slow.
  \end{minipage}
\end{table}

\subsubsection{Scalability of CCL-D}
\label{ssec:scale-test}

Table \ref{table:before-CCL-D} demonstrates strong scalability of CCL-D in diagnosing slow/hang anomalies across heterogeneous clusters of varying sizes. Within two months, we collected diagnostic logs from training tasks spanning tens to 4,000 GPUs to comprehensively evaluate its performance. The distribution of abnormal operations detected in Table \ref{table:before-CCL-D} is as follows: there are 16 cases in the range [16, 128], 9 cases in (128, 512], 4 cases in (512, 1024], 4 cases in (1024, 2048], and 3 cases in (2048, 4000].

Before CCL-D integration, hang detection relied mainly on PyTorch watchdog, which only triggered after the 30-minute timeout. Locating the hang root causes across different GPU scales took an average of 47 hours per successful case. Slow anomalies depended on unpredictable customer feedback and human intuition, with an average location time of 74 hours. Furthermore, coverage of root causes was limited and relied heavily on manual expertise and coarse-grained inference, leading to inefficiency and lack of systematic precision.

With CCL-D, diagnosis becomes automated, covering both intra- and inter-node causes, and anomaly captures increase from 8 to 28. Due to our system's configuration, the system detects hang anomalies within 5 minutes and slow anomalies within 1 minute, and can infer root causes and pinpoint faulty ranks within 1 minute even at the 4,000-GPU scale. Real-world deployments confirm CCL-D as an efficient and accurate solution for large-scale model training, offering a reliable reference for addressing slow/hang anomalies.

\subsection{Anomaly-Free Overhead Measurement}

Diagnostic runtime overhead is evaluated from three perspectives. First, we examine system resource usage, including memory and compute cores. Second, we measure the overhead introduced at the communication operation level. Third, we assess the impact on model training in terms of efficiency and accuracy. For comparison, we select RAS and C4D as baselines, since they operate at the CCL level like CCL-D and do not interrupt model training.

\subsubsection{Overhead of Tracing Framework}



\begin{figure}[t]
  \centering  \includegraphics[width=\linewidth]{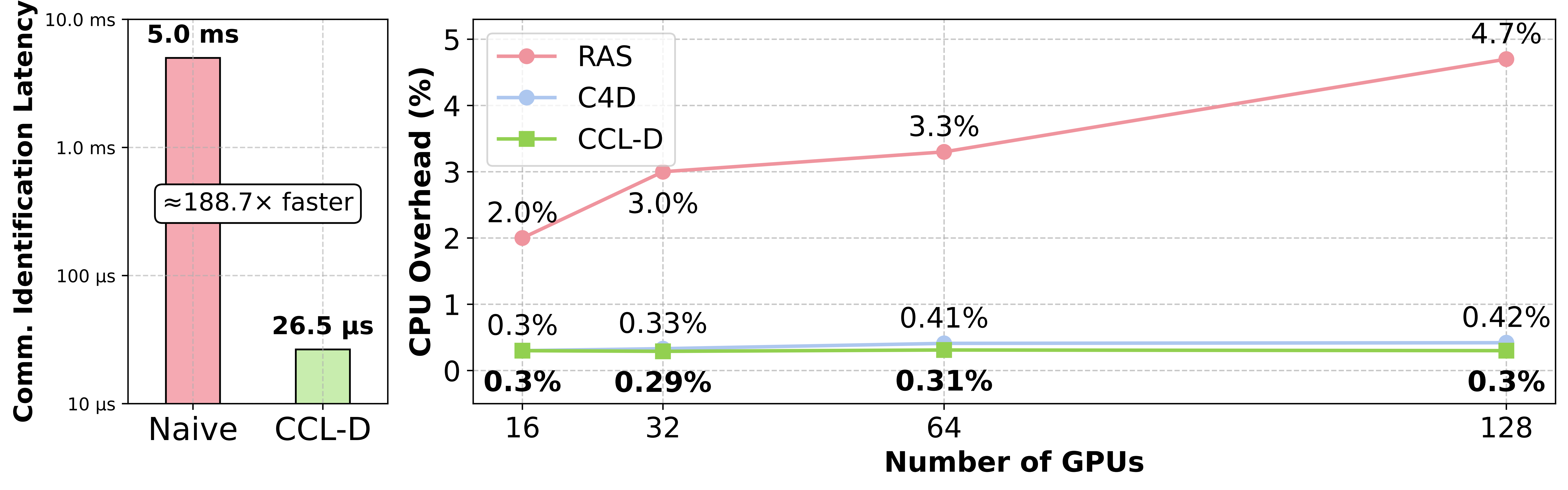}
  \caption{Communication traffic identification overhead and CPU usage per node for anomaly diagnosis at different GPU scales.}
  \label{fig:cpu_usage}
\end{figure}

In this experiment, the differences across parallel strategies of model training are minimal. For clarity, we present results of BaiLing-80B under 16 to 128 GPU configurations with 3D parallel. In terms of resource consumption, RAS, C4D, and CCL-D mainly rely on host-side measurements of collective communication states, incurring negligible overhead on GPU memory and computation. For CPU resources, as shown in the right subfigure of Figure \ref{fig:cpu_usage}, CCL-D maintains a stable CPU utilization of around 0.3\% per node across all scales. C4D shows a similar CPU usage trend as CCL-D. In contrast, RAS overhead increases significantly with the number of ranks, revealing poor scalability.

For host memory usage, CCL-D adopts fixed-size buffers and memory reuse, keeping the per-rank footprint stable at 1184 Bytes regardless of cluster scale. This design aligns with Section \ref{ssec:data_structure}, where the probing frame consists of a 32-byte shared header and a 1152-byte body divided into 8 sub-blocks, each corresponding to one Trace ID and mapped to 8 communication channels. Without using the reserved bytes, each Trace ID occupies 16 Bytes. The memory usage of RAS and C4D could not be accurately measured due to interference from training workloads. Nevertheless, RAS inherently incurs higher memory usage from constructing communicator topologies, and excessive overhead may even disrupt normal collective communication behavior.


Our diagnostic data structures not only incur negligible memory overhead, but also deliver substantial performance gains. To validate their effectiveness, we implemented a centralized naive baseline, where each communication requires a request to the identifier device. As shown in Figure \ref{fig:cpu_usage}, CCL-D reduces once identification latency from milliseconds to nanoseconds, achieving about 188× improvement.


\subsubsection{Communication Operation Measurements}

\begin{figure}[t]
  \centering  
  \includegraphics[width=\linewidth]{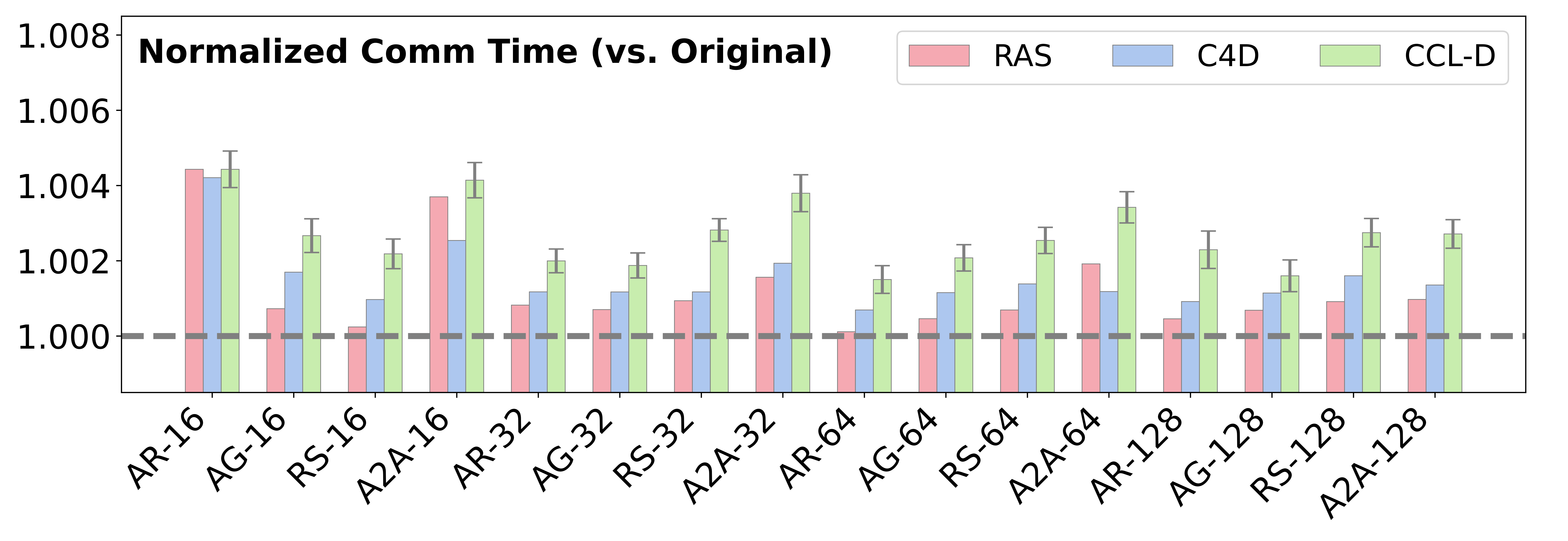}
  \caption{Comparison of normalized communication time (vs. Original) for different operations on 16–128 GPUs.}
  \label{fig:operation_time}
\end{figure}

\begin{figure*}[t]
    \centering
    \begin{subfigure}[b]{0.245\linewidth}
        \includegraphics[width=\linewidth]{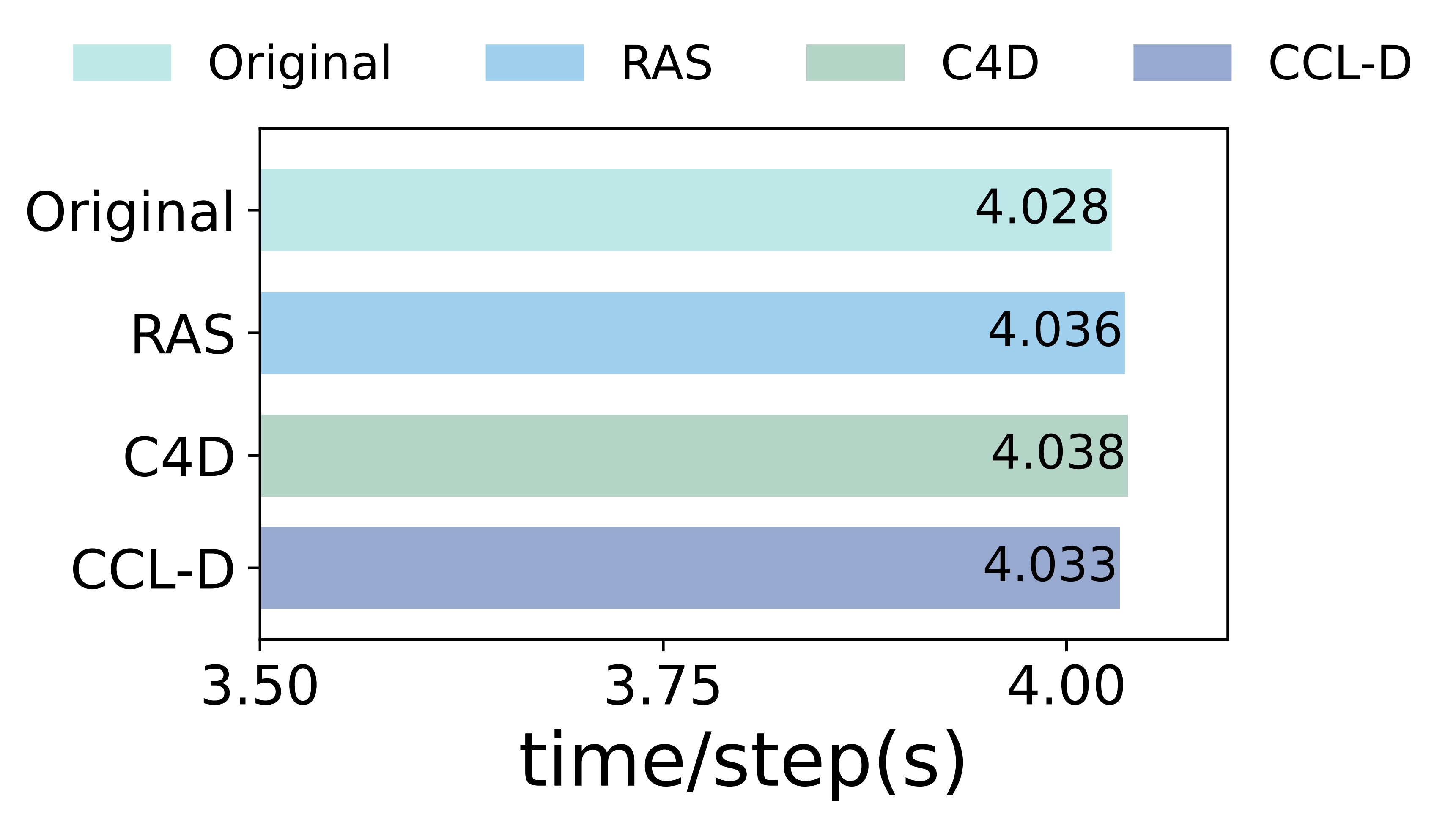}
        \caption{Llama2 per-step time (FSDP)}
        \label{fig:llama2-step}
    \end{subfigure}
    \hfill
    \begin{subfigure}[b]{0.245\linewidth}
        \includegraphics[width=\linewidth]{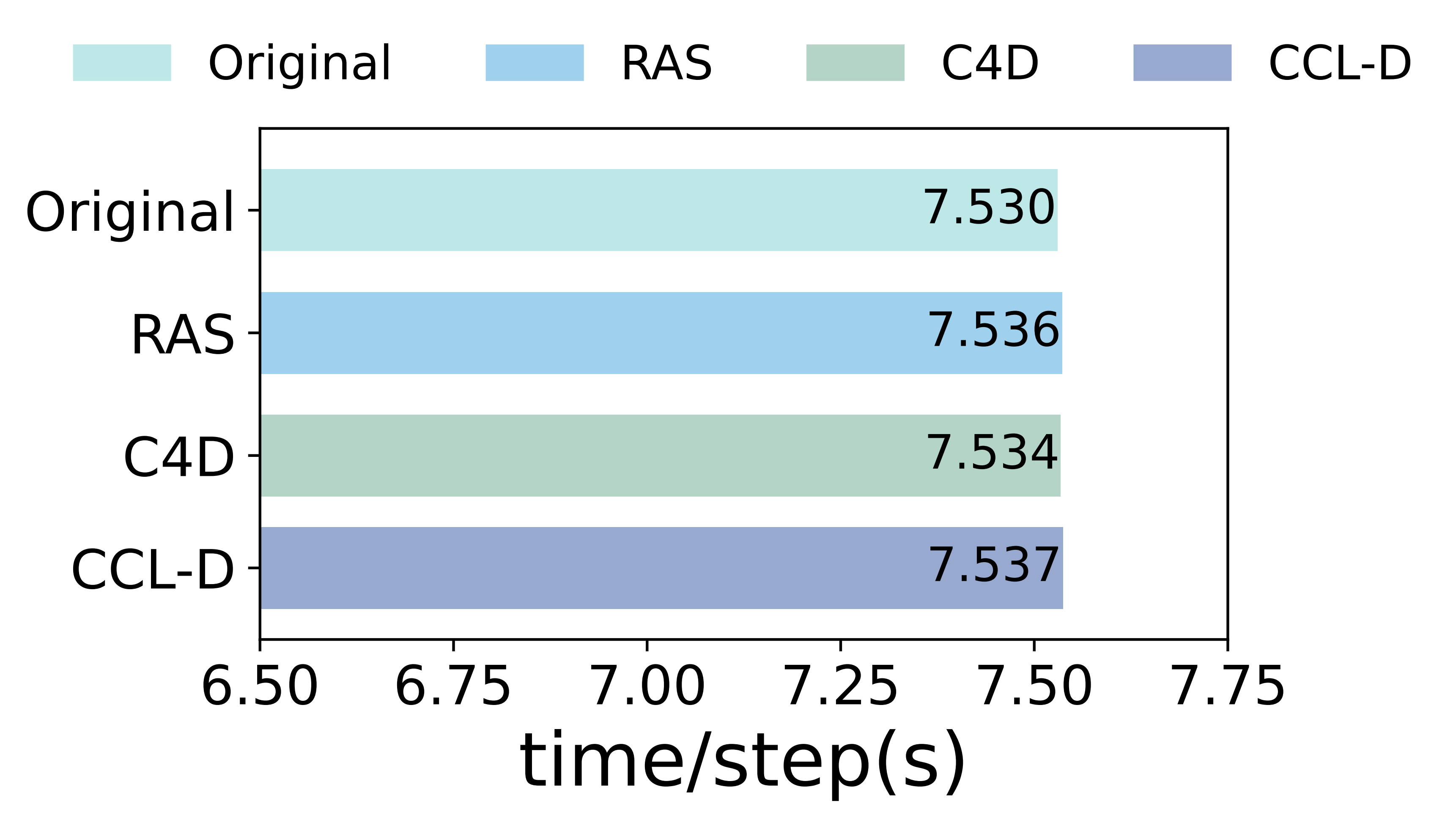}
        \caption{BaiLing-5B per-step time (FSDP)}
        \label{fig:bailing-step}
    \end{subfigure}
    \hfill
    \begin{subfigure}[b]{0.245\linewidth}
        \includegraphics[width=\linewidth]{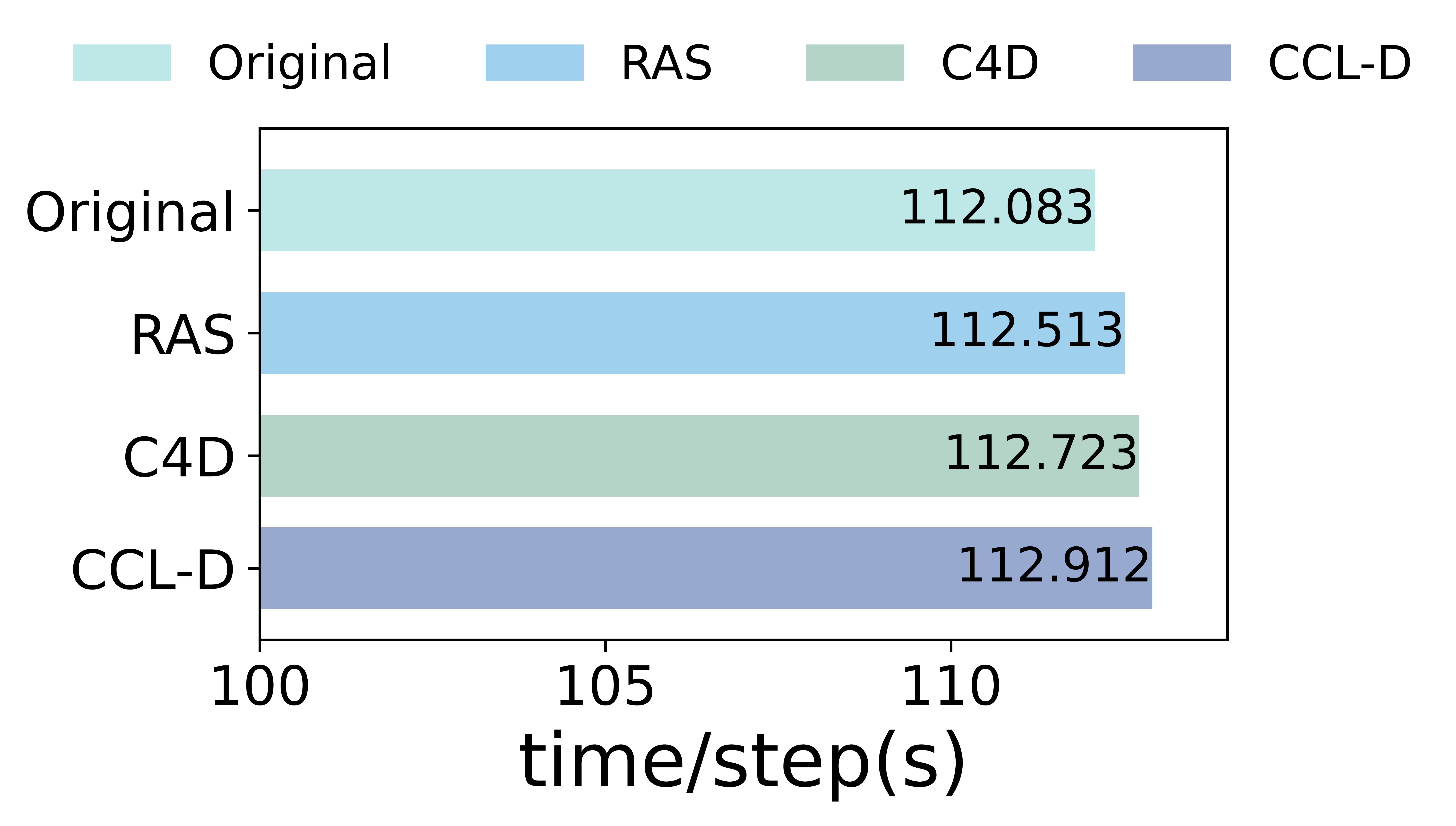}
        \caption{Llama3.1 per-step time (3D)}
        \label{fig:llama3-step}
    \end{subfigure}
    \hfill
    \begin{subfigure}[b]{0.245\linewidth}
        \includegraphics[width=\linewidth]{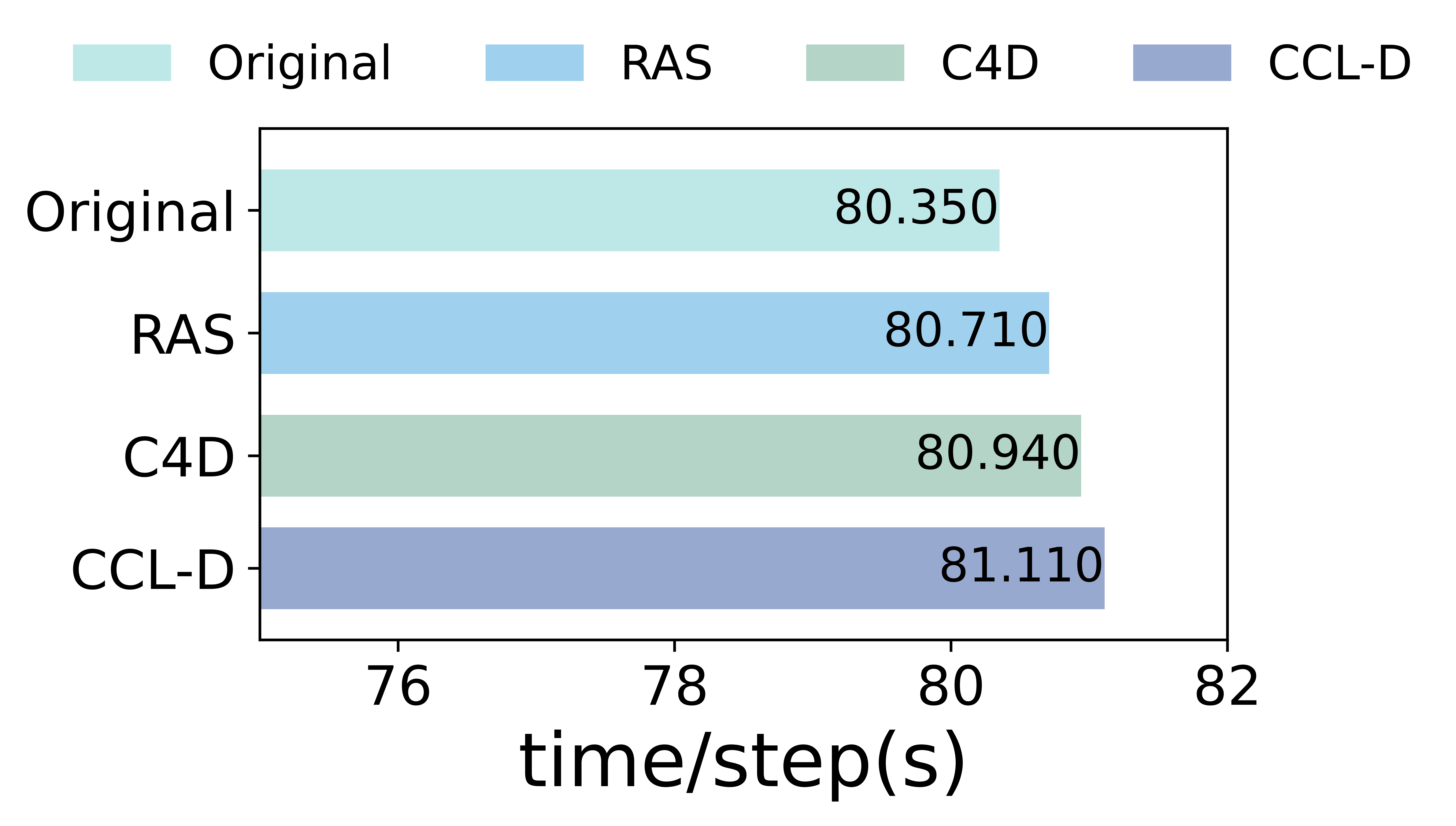}
        \caption{BaiLing-80B per-step time (3D)}
        \label{fig:bailing-80-step}
    \end{subfigure}

    \vspace{0.2cm}
    \begin{subfigure}[b]{0.245\linewidth}
        \includegraphics[width=\linewidth]{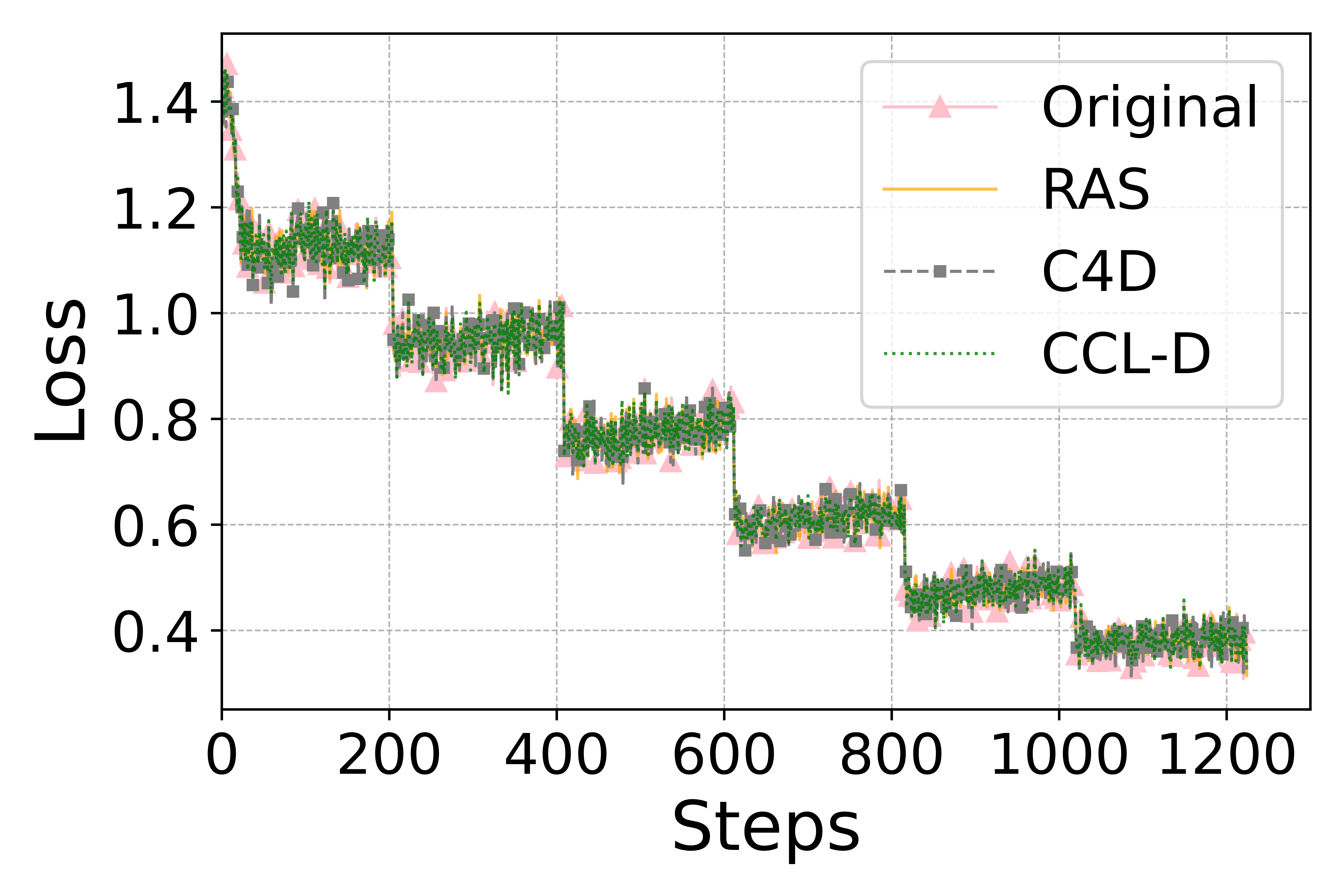}
        \caption{Llama2 loss (FSDP)}
        \label{fig:llama2}
    \end{subfigure}
    \hfill
    \begin{subfigure}[b]{0.245\linewidth}
        \includegraphics[width=\linewidth]{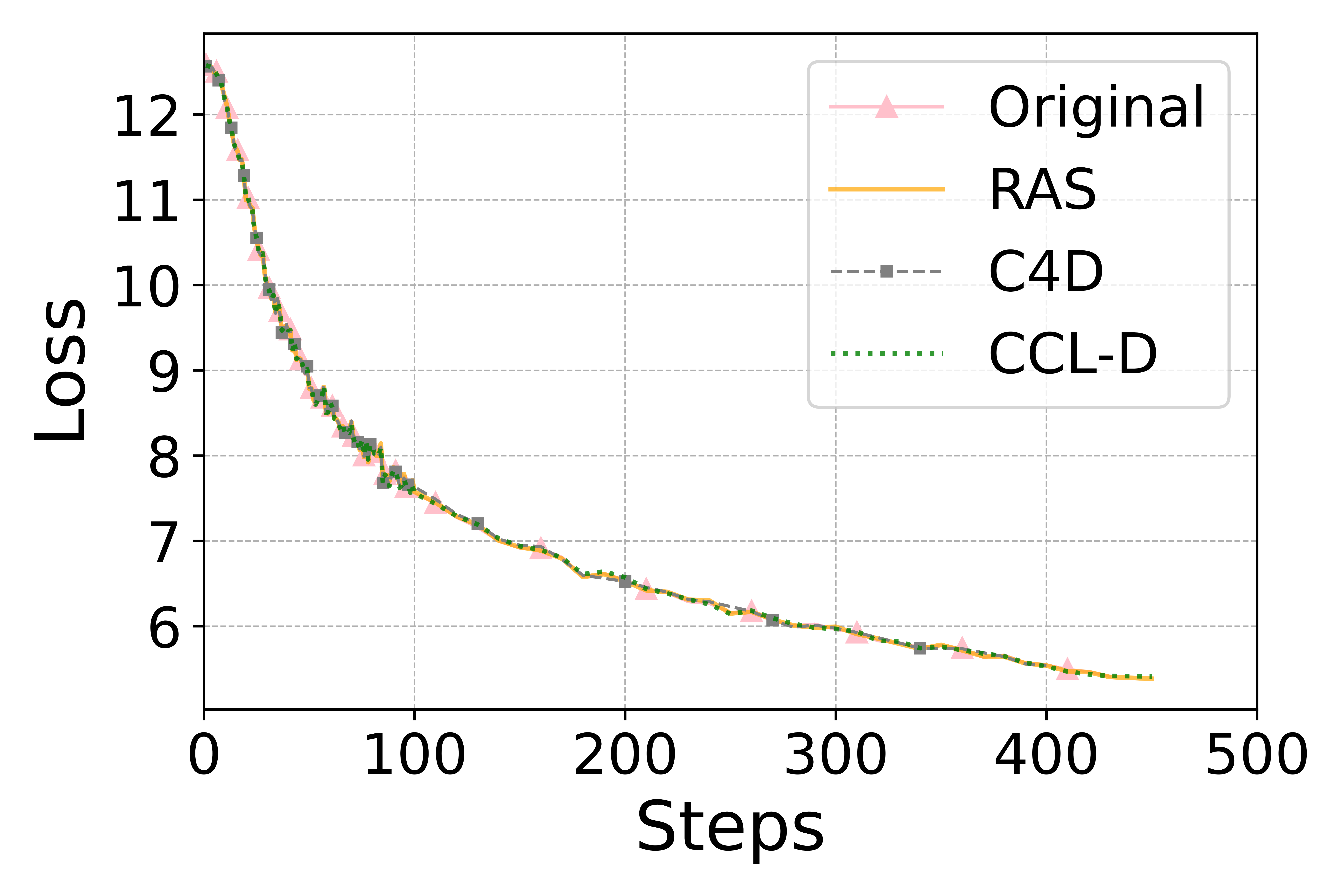}
        \caption{BaiLing-5B loss (FSDP)}
        \label{fig:bailing}
    \end{subfigure}
    \hfill
    \begin{subfigure}[b]{0.245\linewidth}
        \includegraphics[width=\linewidth]{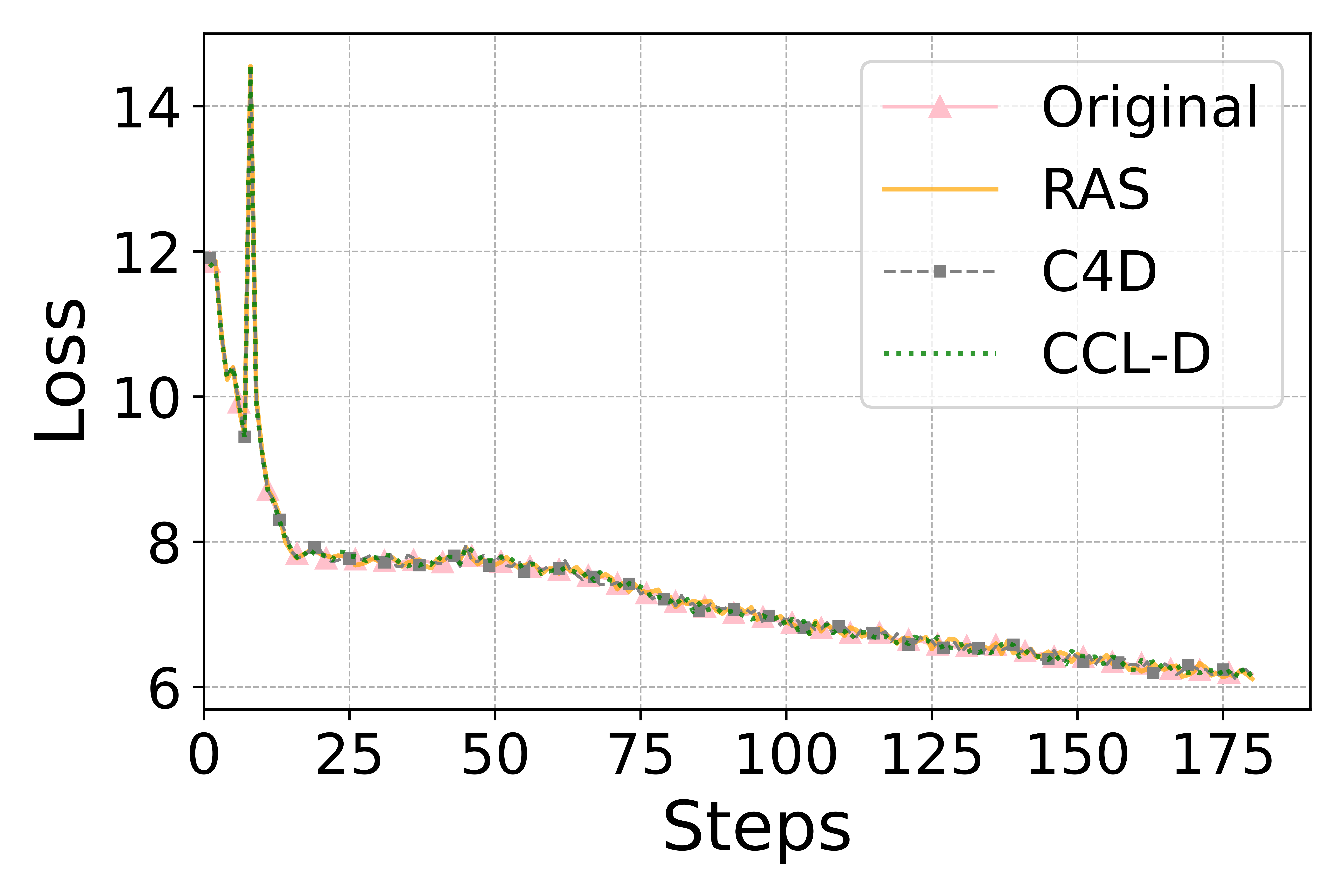}
        \caption{Llama3.1 loss (3D)}
        \label{fig:llama3}
    \end{subfigure}
    \hfill
    \begin{subfigure}[b]{0.245\linewidth}
        \includegraphics[width=\linewidth]{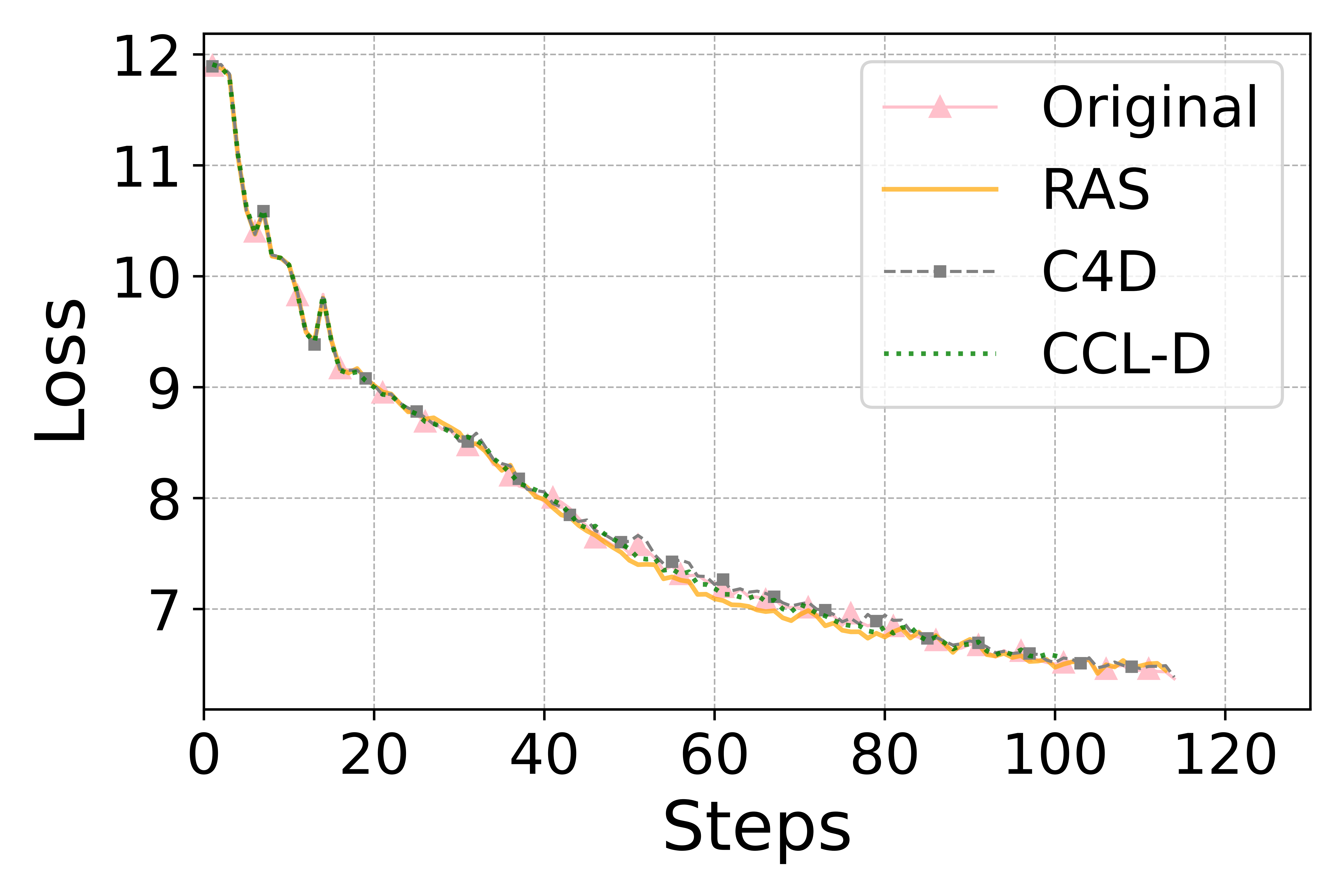}
        \caption{BaiLing-80B loss (3D)}
        \label{fig:bailing-80B}
    \end{subfigure}
    \caption{Per-step time and loss over time under different large models.}
    \label{fig:loss}
    \vspace{-3mm}
\end{figure*}


We evaluate four commonly used collective communication operations in large-scale model training—AllReduce, AllGather, ReduceScatter, and AlltoAll—with data sizes of 64MB, 256MB, 1GB, and 8GB. Each operation is tested 300 times, and the average is reported. Figure \ref{fig:operation_time} presents the normalized communication time with diagnostics relative to the original baseline under 256MB, where the dashed line denotes the original communication. The bar-top segments of CCL-D indicate the variance across different data sizes.

RAS, C4D, and CCL-D all introduce minimal impact on communication operations, with less than 0.45\% additional cost compared to orignal. This low overhead is primarily due to all systems relying on host-driven measurement, which avoids interfering with device-side communication execution. However, CCL-D incurs slightly higher overhead than others, as it measures fine-grained metrics inside GPU kernels to ensure higher diagnostic accuracy. Although RAS achieves lower overhead by only recording operation counts, its limited anomaly coverage is unacceptable.



\subsubsection{Training Efficiency and Accuracy}

We test 4 models with FSDP or 3D parallelism. Llama2-7B and Llama3.1-8B are trained on 16 GPUs, BaiLing-5B on 32 GPUs, and BaiLing-80B on 48 GPUs. Figure \ref{fig:loss} shows the impact of different diagnostic solutions on the training time and loss per step. 


As shown in Figure \ref{fig:loss} (a)–(d), for Llama2 and BaiLing-5B, CCL-D exhibits insensitivity to model types under the same parallel strategy, with maximum overhead as low as 0.12\%. However, compared to FSDP, 3D parallelism offers finer granularity for both training data and model partitioning, resulting in more communicators and higher communication frequency. This lead to 0.74\% and 0.95\% overhead for Llama3.1 and BaiLing-80B. In contrast, RAS, due to its simplistic metric collection, shows lower sensitivity across models and parallel modes, with a maximum overhead of 0.45\%. Nevertheless, reducing runtime overhead at the cost of diagnostic accuracy is not a desirable trade-off.

Figure \ref{fig:loss} (e)-(h) show the training accuracy of different large models. After applying the diagnostic systems, the downward trend of the loss function remains consistent with that of the original training. This is expected, as RAS, C4D, and CCL-D did not modify the model algorithm or training data, thus having no impact on the loss function.

\section{Related Work}


Diagnostic systems are vital for ensuring hardware stability and correct software execution in large-scale training, and have been widely studied across academia and industry.

\textbf{Training-centric anomaly diagnosis systems.}
MegaScale~\cite{jiang2024megascale} scales LLM training to over 10,000 GPUs, employing heartbeat checks, lightweight self-tests, and CUDA event analysis to enhance cluster stability, but its diagnosis remains at the job/node granularity and does not extend into communication kernels. Minder~\cite{deng2025minder} leverages metric similarity and continuity for node-level anomaly identification, showing effectiveness in large production clusters but remaining machine-centric. Dynolog~\cite{Dynolog} integrates PyTorch Profiler into production to capture coarse-grained GPU kernel traces for online monitoring. Superbench~\cite{xiong2024superbench} provides a comprehensive benchmarking suite covering operator, bandwidth, and latency metrics, yet focuses on pre-deployment checks rather than runtime diagnosis. SkeletonHunter~\cite{liu2025skeletonhunter} exploits tensor skeleton features to detect faulty network components, demonstrating strong results under multi-model training tasks. Overall, these approaches enhance observability at system or node granularity but fall short of probing communication kernels, limiting their coverage of slow/hang.


\textbf{General network failure diagnosis systems.}
ComScribe~\cite{akhtar2020comscribe,unat2022monitoring} monitors collective communication using transfer matrices of size and frequency to identify bottlenecks and risks. NetBouncer~\cite{tan2019netbouncer} combines IP-tunnel probing with inference algorithms to detect link and device failures via machine learning. Hostping~\cite{liu2023hostping} and Justitia~\cite{zhang2022justitia} monitor host, bus, and NIC states through hardware sensors and loopback tests, diagnosing anomalies in latency and bandwidth. In contrast, CCL-D operates at the CCL level with lightweight tracing, enabling precise slow/hang diagnosis without external hardware dependency. Moreover, CCL-D is orthogonal to these methods: while network/system-level tools capture infrastructure issues, CCL-D focuses on kernel-level communication states. Their combination offers a more complete diagnostic coverage for large-scale training.

\section{Conclusion}


Slow/Hang anomalies in CCL frequently occur and are difficult to diagnose during large-scale training. In this paper, we introduce CCL-D, which carefully designs a set of high-precision metrics for runtime probing, traces each collective communication in a decentralized manner, and measures the communication with low overhead. Then, it rapidly analyzes the metrics and locates specific faulty ranks using an efficient decision algorithm. Experiments show that CCL-D achieves superior coverage and accuracy, diagnosing slow/hang anomalies in large-scale training within 6 minutes.



\begin{acks}
We would like to acknowledge support from the National Key Research and Development Program of China (Grant No. 2025YFB3003702), the National Natural Science Foundation of China (Grant Nos. 62032023 and T2125013), and the Ant Group SCT. The Al-driven experiments, simulations and model training were performed on the robotic Al-Scientist platform of Chinese Academy of Sciences.
\end{acks}

\bibliographystyle{ACM-Reference-Format}
\bibliography{reference}

\end{document}